\documentclass[%
 reprint,
 superscriptaddress,
 amsmath,amssymb,
 aps,
 prx,
]{revtex4-1}

\usepackage[english]{babel}
\usepackage{graphicx}
\usepackage{dcolumn}
\usepackage{bm}
\usepackage{xcolor}
\usepackage{bbm}
\usepackage[normalem]{ulem}
\usepackage[colorlinks,urlcolor=blue,citecolor=blue,linkcolor=blue]{hyperref}

\begin{document}

\preprint{APS/123-QED}

\title{Homogenisation Theory of Space-Time Metamaterials}

\author{P. A. Huidobro}
 \email{p.arroyo-huidobro@lx.it.pt}
\affiliation{Instituto de Telecomunica\c c\~oes, Instituto Superior Tecnico-University of Lisbon, Avenida Rovisco Pais 1, Lisboa, 1049‐001 Portugal
}
\author{M.G. Silveirinha}%
\affiliation{Instituto de Telecomunica\c c\~oes, Instituto Superior Tecnico-University of Lisbon, Avenida Rovisco Pais 1, Lisboa, 1049‐001 Portugal
}%

\author{E. Galiffi}
\affiliation{The Blackett Laboratory, Department of Physics, Imperial College London, London, SW7 2AZ UK}%
\author{J.B. Pendry}
\affiliation{The Blackett Laboratory, Department of Physics, Imperial College London, London, SW7 2AZ UK}%

\date{\today}

\begin{abstract}
We present a general framework for the homogenisation theory of space-time metamaterials. By mapping to a frame co-moving with the space-time modulation, we derive analytical formulae for the effective material parameters for travelling wave modulations in the low frequency limit: electric permittivity, magnetic permeability and magnetoelectric coupling. Remarkably, we show that the theory is exact at all frequencies in the absence of back-reflections, and exact at low frequencies when that condition is relaxed. This allows us to derive exact formulae for the Fresnel drag experienced by light travelling through travelling-wave modulations of electromagnetic media.
\end{abstract}

\maketitle


\section{\label{sec:level1}Introduction }
Enabled by the advent of new materials and techniques to achieve fast and efficient dynamical modulation of material parameters \cite{shaltout2019spatiotemporal,alam2018large,lira2012electrically}, the emergence of time as a new degree of freedom for the design of metamaterials has recently opened new and intriguing avenues for wave control \cite{caloz2020spacetime}. 
Modulations of a material parameter in time, as well as in space, enable frequency-momentum transitions \cite{interbandWinn1999,lira2012electrically},
non-reciprocal effects \cite{biancalana2007dynamics,yu2009complete,sounas2017non,hadad2016breaking,Taravati2018,Torrent2018,Huidobro2019,camacho2020diffusion}, compact photonic isolators and circulators without magnetic bias \cite{Wang2013Diode, sounas2013giant, Fang2012AharonovBohm}, harmonic generation \cite{chamanara2018linear}, unidirectional amplification \cite{galiffi2019}, topological phases \cite{lin2016photonic,fleury2016floquet,he2019floquet,lustig2018topological} and multifunctional non-reciprocal metasurfaces \cite{Wang2020DiazRubio}. 

Periodic space-time modulations of the permittivity and permeability in space and time following a travelling-wave form, 
\begin{align}
    \epsilon(x,t) &= \epsilon(x-vt), &
    \mu(x,t)      &= \mu(x-vt) ,
\end{align}
have attracted much attention since early research \cite{oliner1961wave,cassedy1963dispersion,cassedy1967dispersion}. In these expressions, $v$ stands for the modulation speed, which, since we are concerned with modulations and not with moving media, is not bounded by the speed of light. Figure 1(a) shows a sketch of a sinusoidal space-time modulations. The spatial, $g$, and temporal, $\Omega$, modulation frequencies, determine the modulation speed as $v=\Omega/g$. Travelling wave modulations impose a linear bias, breaking time-reversal symmetry and resulting in non-symmetric high frequency band gaps, which can be exploited for frequency-momentum transitions and nonreciprocal devices (see Fig. 1c) \cite{lira2012electrically,sounas2017non}. Recently, it has been shown that the need for working at high (band-gap) frequencies can be lifted and non-reciprocity emerges as a linear broad-band phenomenon in \textit{luminal} modulations of the permittivity. Space-time modulations at speeds approaching that of waves in a medium result in non-reciprocal broadband amplification \cite{galiffi2019}. Interestingly,  nonreciprocity can be achieved in the long-wavelength limit and even at zero frequency if both electromagnetic parameters, $\epsilon$ and $\mu$, are modulated \cite{Taravati2018} (see Fig. 1d), realising a synthetic, tunable form of Fresnel drag \cite{Huidobro2019}. 
\begin{figure}
    \centering
    \includegraphics{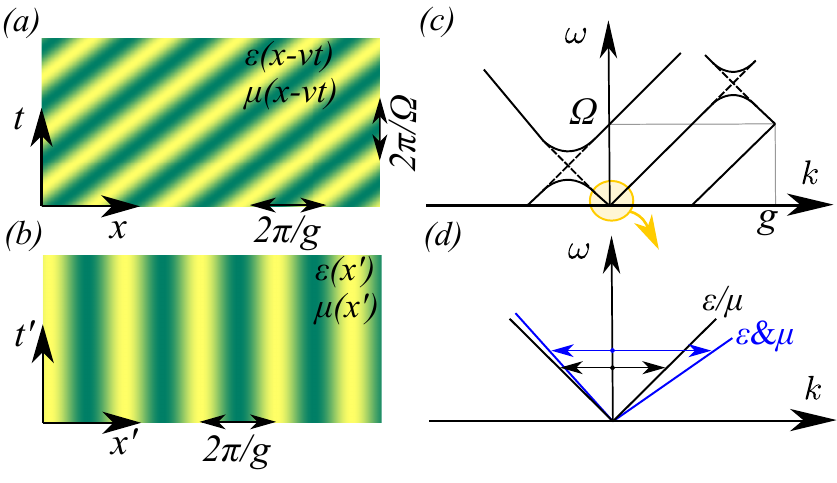}
    \caption{Space-time modulated metamaterials. (a,b) Travelling-wave modulations of the electromagnetic parameters as seen from the laboratory frame (a, unprimed coordinates) and from a frame co-moving with the modulations (b, primed coordinates). (c) Sketch of a representative band diagram of space-time modulated media: the bands are displaced by the space-time reciprocal lattice vector (g, $\Omega$). Non-symmetric band gaps open in non impedance-matched systems. (d) In the long wavelength limit the response is nonreciprocal only if both the permittivity and the permeability are modulated in space and time. }
    \label{fig:sketch}
\end{figure}{}

Despite the broad interest raised by space-time media, most of the theoretical tools employed for their analysis are based on semi-analytical or numerical approaches, such as Floquet-Bloch theory \cite{cassedy1963dispersion,cassedy1967dispersion}, transfer matrix \cite{LiCummer2019}, finite element methods \cite{taravati2017nonreciprocal}, or perturbative Floquet-Bloch approaches \cite{Torrent2018,Huidobro2019}. Here we present the first analytical theory of space-time electromagnetic metamaterials. By transforming Maxwell's equations to the frame co-moving with the modulation, we develop a homogenisation theory, deriving closed-form expressions to calculate the effective electromagnetic parameters of space-time modulated media. This allows us to formulate an effective medium description, which, remarkably, we show is exact at any frequency in the absence of back-reflections, (that is, if the system is impedance-matched), and in the metamaterial (long-wavelength) limit for the impedance-mismatched case. We show that our analytical formalism can be applied to stratified or sinusoidal travelling-wave modulations as long as a Bloch wave picture is valid, enabling the identification and characterization of different regimes of non-reciprocity in spacetime-modulated media.

\section{Homogenisation theory}
The fields in space-time modulated media 
 satisfy Maxwell's equations,
\begin{align}
    \label{eq:MaxwellLab}
    \nabla\times\mathbf{E} &= -\frac{\partial \mathbf{B}}{\partial t}, &
    \nabla\times\mathbf{H} &= \frac{\partial \mathbf{D}}{\partial t},
\end{align}
and are related through the constitutive equations as follows, 
\begin{eqnarray}
    \mathbf{D}(x,y,z,t) &=& \epsilon(x-vt)  \mathbf{E}(x,y,z,t), \\
    \mathbf{B}(x,y,z,t) &=& \mu(x-vt)  \mathbf{H}(x,y,z,t), 
\end{eqnarray}
where $\epsilon$ and $\mu$ are deemed to include $\epsilon_0$ and $\mu_0$, respectively, which determine the speed of light in vacuum as $c_0=1/\sqrt{\epsilon_0\mu_0}$.
We note that this represents the spatio-temporal modulation along one direction of otherwise isotropic but possibly inhomogeneous permittivity and permeability. Here we assume the system is not dispersive although it is possible to generalize the theory to include dispersion.    

Let us now consider a Galilean transformation to a co-moving frame ($x'=x-v t$, $y'=y$, $z'=z$, $t'=t$). We have, for the parallel component of the fields, 
\begin{align} \label{eq:parallelDB}
    \mathbf{D'}_{||} &= \epsilon\left(x'\right)  \mathbf{E'}_{||}, &
     \mathbf{B'}_{||} &= \mu\left(x'\right)  \mathbf{H'}_{||},
\end{align}{}
where the primed fields depend on the transformed coordinates, $(x',y',z',t') $.
The perpendicular components are transformed as (see S.M.), 
\begin{eqnarray}
    \label{eq:perptransformed}
    \begin{bmatrix}{}
        \mathbf{D'}_{\perp} \\
         \mathbf{B'}_{\perp}
    \end{bmatrix} &=& \frac{1}{1-\epsilon(x')\mu(x') v^2} \times \\
    &&\begin{bmatrix}
        \epsilon(x') \mathbbm{1}& -\epsilon(x')\mu(x')\mathbf{v}\times\mathbbm{1}\\ 
        \epsilon(x')\mu(x')\mathbf{v}\times\mathbbm{1} & \mu(x') \mathbbm{1} 
    \end{bmatrix}
    \begin{bmatrix}{}
        \mathbf{E'}_{\perp} \\
         \mathbf{H'}_{\perp}
    \end{bmatrix} \nonumber
\end{eqnarray}{}
where $\mathbf{v}$ is the space-time modulation velocity vector, and $ \mathbbm{1}$ is the $2\times 2$ identity matrix.  This shows that in the co-moving frame, the modulation of parameters in space and time results in a moving-medium type coupling between the electric and magnetic fields. Hence, a bianisotropic coupling arises and the electromagnetic response is nonreciprocal. Interestingly, this is different from the usual moving medium situation, where the bianisotropic coupling arises in the lab-frame while in the co-moving frame all interactions are reciprocal. Hence, from Eqs. (7-9) we can write the effective constitutive parameters of the space-time modulated media in the co-moving frame as,
\begin{eqnarray}
    \begin{bmatrix}
        \mathbf{D}' \\   \mathbf{B}'
    \end{bmatrix}    = 
    \begin{bmatrix}
        \hat{\boldsymbol{\epsilon}}' & \hat{\boldsymbol{\xi}}' \\ \hat{\boldsymbol{\zeta}}'  &   \hat{\boldsymbol{\mu}}'
    \end{bmatrix}      
    \begin{bmatrix}
        \mathbf{E}' \\   \mathbf{H}'
    \end{bmatrix}  
\end{eqnarray}
as those representing an uniaxial medium, 
\begin{eqnarray}
\label{eq:epsmatrixcomov}
    \hat{\boldsymbol{\epsilon}}' &=&
    \begin{bmatrix}
         \epsilon_{||}' & 0 & 0 \\ 
         0 & \epsilon_{\perp}' & 0 \\
        0 & 0 & \epsilon_{\perp}'
    \end{bmatrix}, \,
        \hat{\boldsymbol{\mu}}' = 
    \begin{bmatrix}
         \mu_{||}' & 0 & 0 \\ 
         0 & \mu_{\perp}' & 0 \\
        0 & 0 & \mu_{\perp}'
    \end{bmatrix} ,
\end{eqnarray}
with a nonreciprocal magnetoelectric coupling $\hat{\boldsymbol{\zeta}}'=-\hat{\boldsymbol{\xi}}'$, and
\begin{eqnarray}
\label{eq:ximatrixcomov}
    \hat{\boldsymbol{\xi}}' &=&
    \begin{bmatrix}
         0 & 0 & 0 \\ 
         0 & 0 & \xi' \\
        0 & -\xi & 0
    \end{bmatrix},
\end{eqnarray}
with
\begin{eqnarray}
    \label{eq:epscomov}
    \epsilon'_{||}(x') &=& \epsilon(x'), \\ 
    \mu'_{||}(x') &=& \mu(x'), \\ 
    \epsilon'_{\perp}(x') &=&  \frac{\epsilon(x')}{1-\epsilon(x')\mu(x') v^2} , \\ 
    \mu'_{\perp}(x') &=&  \frac{\mu(x')}{1-\epsilon(x')\mu(x') v^2}  , \\ 
    \label{eq:xicomov}
    \xi'(x') &=& -v  \frac{\epsilon(x')\mu(x')}{1-\epsilon(x')\mu(x') v^2}  ,
\end{eqnarray}{}
In this frame, all the quantities in the constitutive matrix depend solely on $x'$, so we can write the effective parameters by homogenising over the unit cell following the conventional procedure. From the continuity of the normal components of $\mathbf{D}$ and $\mathbf{B}$ at an interface, we have that in the long wavelength limit, $\langle\mathbf{D}'_{||}\rangle = \mathbf{D}'_{||}$ and $\langle\mathbf{B}'_{||}\rangle = \mathbf{B}'_{||}$. Then, for the parallel components of the fields, we write Eqs. \ref{eq:parallelDB} as $ \langle \mathbf{D}'_{||}/\epsilon(x')\rangle =  \langle \mathbf{E}'_{||} \rangle $ and $ \langle \mathbf{B}'_{||}/\mu(x')\rangle =  \langle \mathbf{H}'_{||} \rangle $, and we have,
\begin{eqnarray}
  \langle \mathbf{D}'_{||} \rangle &=&  \Big\langle \frac{1}{\epsilon(x')} \Big\rangle^{-1}  \langle \mathbf{E}'_{||} \rangle, \\
    \langle \mathbf{B}'_{||} \rangle &=&  \Big\langle \frac{1}{\mu(x')} \Big\rangle^{-1}  \langle \mathbf{H}'_{||} \rangle.
\end{eqnarray}
Hence, the the effective permittivity and permeability in the parallel direction are given by,
\begin{eqnarray}
    \label{eq:epseff}
    \overline{\epsilon'}_{||} &=& \left[ \frac{1}{d} \int_0^d\frac{1}{\epsilon'_{||}(x')}\,\text{d}x'\right]^{-1} \\ 
    \label{eq:mueff}
     \overline{\mu'}_{||} &=& \left[ \frac{1}{d} \int_0^d\frac{1}{\mu'_{||}(x')}\,\text{d}x'\right]^{-1} ,
\end{eqnarray} 
where $d$ is the spatial periodicity of the modulation. On the other hand, from the continuity of the tangential components of $\mathbf{E}$ and $\mathbf{H}$ at an interface, we have that $\langle\mathbf{E}'_{\perp}\rangle = \mathbf{E}'_{\perp}$ and $\langle\mathbf{H}'_{\perp}\rangle = \mathbf{H}'_{\perp}$. 
Hence, for the perpendicular components of the fields, we have from Eqs. \ref{eq:perptransformed}, $\langle\mathbf{D}'_\perp\rangle = \langle \boldsymbol{\epsilon}_\perp(x') \mathbf{E} + \boldsymbol{\xi}(x') \mathbf{H}' \rangle $ and $\langle\mathbf{B}'_\perp\rangle = \langle \boldsymbol{\mu}_\perp(x') \mathbf{H} - \boldsymbol{\xi}(x') \mathbf{E}' \rangle $, and,
\begin{eqnarray}
    \langle\mathbf{D}'_\perp\rangle &=& \langle \boldsymbol{\epsilon}_\perp(x')\rangle \langle\mathbf{E}'_\perp\rangle + \langle\boldsymbol{\xi}(x')\rangle \langle\mathbf{H}'_\perp \rangle \\ \langle\mathbf{B}'_\perp\rangle &=& \langle \boldsymbol{\mu}_\perp(x') \rangle \langle\mathbf{H}'_\perp\rangle - \langle\boldsymbol{\xi}(x')\rangle  \langle\mathbf{E}'_\perp\rangle 
\end{eqnarray}{}
Hence, the remaining effective parameters are given by,
\begin{eqnarray}
\label{eq:epsperpeff}
  \overline{ \epsilon'}_{\perp} &=& \frac{1}{d} \int_0^d  \epsilon'_{\perp}(x') \,\text{d}x' , \\ 
  \label{eq:muperpeff}
    \overline{\mu'}_{\perp} &=& \frac{1}{d} \int_0^d  \mu'_{\perp}(x')\, \text{d}x' , \\ 
    \label{eq:xieff}
    \overline{\xi'} &=&  \frac{1}{d} \int_0^d  \xi'(x')\, \text{d}x' ,
\end{eqnarray}  
Here, we have assumed a long wavelength approximation ($\omega d/c_0 \ll 1$, and $\omega$ much smaller than the temporal modulation frequency). However, this restriction can be lifted in the absence of back-reflections, as we will show below, and thus the expressions are exact at any frequency for impedance-matched systems where $\mu(x,t)/\epsilon(x,t)= Z^2 $ is a constant. 

The above set of equations, \ref{eq:epseff}, \ref{eq:mueff} and \ref{eq:epsperpeff}-\ref{eq:xieff}, provide the effective medium description of space-time modulations of travelling-wave form in the co-moving frame.  The last step is to transform them to the laboratory frame (see S.M.), where the uniaxial and non-reciprocal structure of the parameters is maintained. The effective medium parameters in the stationary frame have the same matrix form as in the co-moving frame, Eqs. \ref{eq:epsmatrixcomov}-\ref{eq:ximatrixcomov}, with components given by
\begin{eqnarray}
\label{eq:effparamlab}
    \epsilon_{||}^\text{eff} &=&  \overline{\epsilon'}_{||} , \\
    \mu_{||}^\text{eff} &=&   \overline{\mu'}_{||}, \\
    \epsilon_{\perp}^\text{eff} &=&  \frac{ \overline{\epsilon'}_\perp} { (1-v  \overline{\xi'})^2 - v^2  \overline{\epsilon}'_\perp \, \overline{\mu}'_\perp } , \\
    \mu_{\perp}^\text{eff} &=&  \frac{ \overline{\mu'}_\perp} { (1-v  \overline{\xi'})^2 - v^2  \overline{\epsilon'}_\perp \, \overline{\mu}'_\perp} , \\ 
    \xi^\text{eff} &=&  -\frac{v \overline{\epsilon'}_\perp\,  \overline{\mu'}_\perp +  (1-v \overline{\xi'})  \overline{\xi'} } { (1-v  \overline{\xi'})^2 -v^2  \overline{\epsilon'}_\perp\,  \overline{\mu'}_\perp }.
    \label{eq:effparamlab2}
\end{eqnarray}{}
This set of equations, together with Eqs. \ref{eq:epseff}, \ref{eq:mueff} and \ref{eq:epsperpeff}-\ref{eq:xieff}, constitute the main result of this paper. They prescribe how to calculate the effective material parameters of any space-time modulation of travelling-wave type. With them, we can write the effective mode dispersion relation as,
\begin{eqnarray}
    \mu^\text{eff}_\perp \epsilon^\text{eff}_\perp \omega^2 = \frac{ \mu^\text{eff}_\perp}{ \mu^\text{eff}_{||}} k_y^2 + (k-\xi^\text{eff}\omega)^2, \, \text{s-polarisation,} \\
     \mu^\text{eff}_\perp \epsilon^\text{eff}_\perp \omega^2 = \frac{ \epsilon^\text{eff}_\perp}{ \epsilon^\text{eff}_{||}} k_y^2 + (k-\xi^\text{eff}\omega)^2, \, \text{p-polarisation,}
\end{eqnarray}
Finally, from this we obtain the effective group velocity in terms of the effective parameters,
\begin{equation}
    \label{eq:effvel}
        v_\text{eff}^{\pm} =\frac{1}{\pm\sqrt{\epsilon_{\perp}^{\text{eff}}\mu_{\perp}^{\text{eff}}} + \xi^{\text{eff}}},
\end{equation}{}
where the $\pm$ sign correspond to forward and backward propagating waves, respectively.

Importantly, our derived formulae show that for any kind of travelling-wave modulations where only one of the parameters is modulated, that is, only the permittivity, or only the permeability, then $\xi^\text{eff}=0$ (see S.M. for a detailed proof), and the effective medium is reciprocal. In this case, $v_\text{eff}^+ = v_\text{eff}^-$, as expected.

\section{Exact theory in the absence of back-scattering}
We now show that the above theory is in fact exact at any frequency in the absence of back-scattering. In matched space-time modulated systems, Maxwell's equations can be solved analytically in the co-moving frame. This is seen by noting that, with equal modulations of the permittivity and the permeability the medium impedance is constant,
\begin{eqnarray}
    \frac{\mu(x,t)}{\epsilon(x,t)}=Z^2,
\end{eqnarray}
and we can write, 
\begin{align}
    \label{eq:impedances}
    E_\perp &=  \pm \sigma_{s,p} Z  H_\perp, &
    D_\perp &=  \pm  \sigma_{s,p} Z^{-1}  B_\perp,
\end{align}
where the top (bottom) sign corresponds to forward (backward) propagating waves, and $\sigma_s=-1$ for s-polarisation ($E_\perp=E_z$), and $\sigma_p=+1$ for p-polarisation ($H_\perp=H_z$). At normal incidence, the perpendicular component of Maxwell's equations in the laboratory frame, Eqs. \ref{eq:MaxwellLab}, read as, 
\begin{align}
    \sigma_{s,p} \partial_x E_\perp &= - \partial_t B_\perp, &
   \sigma_{s,p} \partial_x H_\perp &= - \partial_t D_\perp.
\end{align}
Making use of Eqs. \ref{eq:impedances}, and given that the medium impedance is constant, both equations reduce to one (see S.M. for more details),
\begin{eqnarray}
  \frac{\partial}{\partial x} \left[ \epsilon(x,t)^{-1} D_\perp \right] &=& \mp Z \frac{\partial D_\perp}{\partial t}  ,
\end{eqnarray}
with the $-$ and $+$ signs corresponding to forward and backward wave propagation, respectively. By transforming to the Galilean frame co-moving with the space-time modulation, we arrive to a partial differential equation for $D_\perp$, 
\begin{eqnarray}
  \frac{\partial}{\partial x'} \left[ \left( \pm c(x') -  v \right) D_\perp \right] &=& -  \frac{\partial D_\perp}{\partial t'},
\end{eqnarray}
where we have introduced $c(x')= Z^{-1}\epsilon(x')^{-1}$ as the local wave velocity. In the frequency domain, we arrive at,
\begin{eqnarray}
\label{eq:analyticalfieldPDE}
  \ln{\left[ D_\perp\left( \pm c(x') -  v  \right) \right]} = i\omega' \int  \frac{1}{  \pm c(x') -  v  } \text{d}x'.
\end{eqnarray}
we can identify an effective wave-vector by considering that, given the absence of back-reflections, the phase accumulated in one spatial period equals the phase-shift of Bloch modes across the unit cell, $\Delta\phi = k_\text{eff}^{\pm} d$. Hence, we have 
\begin{eqnarray}
\label{eq:effkpm}
  k^{\pm} = \omega'\frac{1}{d} \int_0^d \frac{1}{  \pm c(x') -  v  } \text{d}x'.
\end{eqnarray}
When the grating modulation speed is large enough to approach the local velocity of light at any point within the grating, there is a singular point at $c(x') = v$, where the integrand diverges for waves that co-propagate with the space-time modulation ($+$ sign in the integrand). As a result, the effective group velocity of forward waves in the co-moving frame approaches zero. At this point, in the laboratory frame, the dispersion lines of forward modes of different orders get arbitrarily close and are parallel to the dispersion of waves in the background (unmodulated) medium. This marks the onset of the previously identified \textit{luminal} regime \cite{cassedy1963dispersion}, where Bloch theory fails and an effective medium description of the system is no longer valid, although our analytical approach can be further extended to the study of this regime \cite{pendry2020new}.

Finally, from the dispersion relation of forward and backward propagating waves in the co-moving frame, $k^{\pm} = \omega (  \pm\sqrt{\epsilon'_\perp\mu'_\perp} - \xi' ) $, we can derive the effective parameters as
\begin{eqnarray}
    \epsilon'_\perp &=& \mu'_\perp = \sqrt{\epsilon'_\perp\mu'_\perp}=  \frac{1}{2\omega} (k^+ - k^-) \nonumber \\ && = \frac{1}{d} \int_0^d \frac{\epsilon(x')}{1-\epsilon(x')\mu(x')v^2} \text{d}x',\\
    \xi' &=& -\frac{1}{2\omega} (k^++k^-) \nonumber \\ 
    && = - v \frac{1}{d} \int_0^d \frac{\epsilon(x')\mu(x')}{1-\epsilon(x')\mu(x')v^2} \text{d}x',
\end{eqnarray}
in agreement with our homogenisation formulae. This proves that the homogenisation theory is in fact exact in the absence of back-scattering. In the S.M. we provide an alternative proof based on transfer matrix theory. 


In the following, we apply our formulae to different cases of travelling-wave media, and write analytical effective medium parameters for the cases of travelling wave stratified and sinusoidal space-time media. 

\begin{figure}
    \centering
    \includegraphics{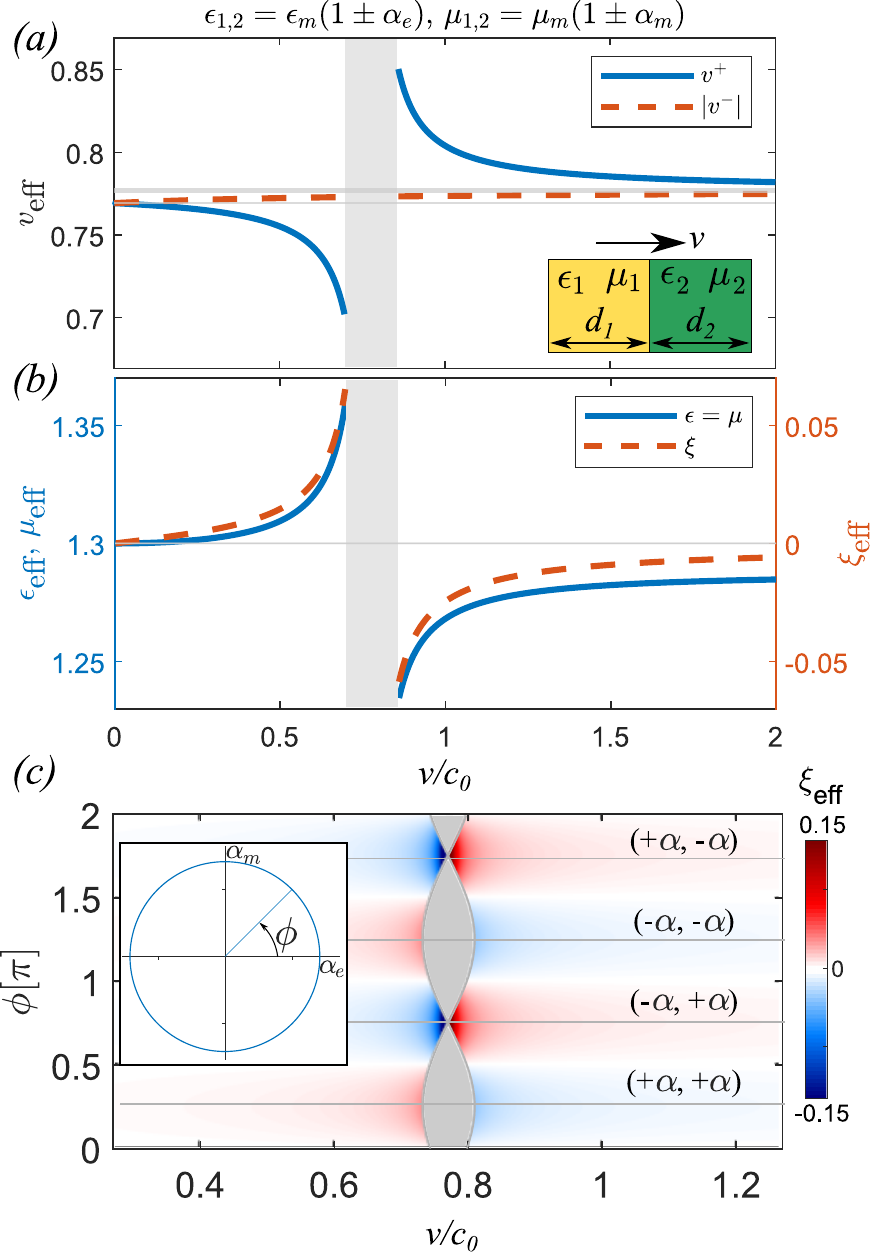}
    \caption{Group velocity and effective parameters for a travelling bi-layered medium as a function of speed, $v$. (a) Group velocity, in absolute value, of forward (blue, solid) and backward (orange, dashed) waves, for matched space-time modulations, $\alpha_e=\alpha_m=0.1$. The range of modulations where a homogenisation picture is not valid, $c_1<v/c_0<c_2$ is marked with a gray area. (b) Effective permittivity, $\epsilon_\perp^\text{eff}$, and permeability, $\mu_\perp^\text{eff}$, (left axis), and magnetoelectric coupling  $\xi_\text{eff}$ (right axis). (c) Effective magnetoelectric coupling as a function of space-time modulation speed, $v$, and the values of $(\alpha_e,\alpha_m)$, parametrized by $\phi$ as shown in the inset panel. We take $\epsilon_m=\mu_m=1.3$ and $d_1=d_2=d/2$. We show relative values of the wave velocity and effective parameters, i.e., $v_\text{eff}$ is in units of $c_0$, $\epsilon_\text{eff} $ and $\mu_\text{eff} $ in units of $\epsilon_0 $ and $\mu_0$, respectively, and $\xi_\text{eff} $ in units of $c_0^{-1}$. }
    \label{fig:stratified}
\end{figure}{}

\section{Travelling stratified media}
Let us consider a travelling two-layer stratified medium with relative parameters $(\epsilon_1,\mu_1)$ and $(\epsilon_2,\mu_2)$, and thicknesses $d_1$ and $d_2$ (period $d=d_1+d_2$) modulated at speed $v$. In this case, the integrals in Eqs. \ref{eq:epscomov}-\ref{eq:xieff} straightforwardly give the set of effective parameters in the co-moving frame, 
\begin{eqnarray}
    \label{eq:epscomovstr}
    \frac{\overline{\epsilon}'_{||}}{\epsilon_0} &=& \left(\frac{d_1}{d}\frac{1}{\epsilon_1}+\frac{d_2}{d}\frac{1}{\epsilon_2}\right)^{-1} , \\ 
     \frac{\overline{\mu}'_{||}}{\mu_0} &=& \left(\frac{d_1}{d}\frac{1}{\mu_1}+\frac{d_2}{d}\frac{1}{\mu_2}\right)^{-1}, \\ 
         \label{eq:epsperpcomovstr}
   \frac{\overline{ \epsilon}'_{\perp}}{\mu_0} &=& \frac{1}{d} \left(\frac{\epsilon_1}{1-\epsilon_1\mu_1v^2/c_0^2}d_1 +\frac{\epsilon_2}{1-\epsilon_2\mu_2v^2/c_0^2}d_2\right)  , \\ 
    \frac{\overline{\mu}'_{\perp}}{\mu_0} &=& \frac{1}{d} \left(\frac{\mu_1}{1-\epsilon_1\mu_1v^2/c_0^2}d_1 +\frac{\mu_2}{1-\epsilon_2\mu_2v^2/c_0^2}d_2\right) , \\ 
    \label{eq:xicomovstr}
    \overline{\xi'}c_0 &=& - \frac{v}{dc_0} \left(\frac{\epsilon_1\mu_1 d_1}{1-\epsilon_1\mu_1v^2/c_0^2} +\frac{\epsilon_2\mu_2 d_2}{1-\epsilon_2\mu_2v^2/c_0^2}\right) ,
\end{eqnarray}{}
which need to be transformed to the rest frame through Eqs. \ref{eq:effparamlab}-\ref{eq:effparamlab2}. From the lab-frame effective parameters, which are given in the S.M., the effective wave velocity can be obtained through Eq. \ref{eq:effvel}. Alternatively, for matched systems, the effective wave velocity can be derived from the exact Eq. \ref{eq:effkpm}, and transforming to the rest frame (see S.M.).

We now particularize to a travelling bilayer crystal with $\epsilon_{1,2}=\epsilon_m(1\pm\alpha_e)$, $\mu_{1,2}=\mu_m(1\pm\alpha_m)$, such that the permittivity and permeability are symmetrically shifted above and below the background values, $\epsilon_m$, $\mu_m$. We first consider the case of matched space-time modulations, with $\alpha_{e,m}=\alpha=0.1$, and constant impedance $Z_1=Z_2=Z_0\sqrt{\mu_m/\epsilon_m}$, and $d_1=d_2$. In this case, the theory is exact and the effective parameters in the rest frame reduce to, 
\begin{eqnarray}
\label{eq:effepsstrat}
 \frac{\epsilon^\text{eff}_{\perp} }{\epsilon_0}&=&    \epsilon_m \frac{1-\left(1- \alpha ^2\right) v^2c_m^{-2}c_0^{-2}}{1-v^2c_m^{-2}c_0^{-2}}
 \\ 
 \frac{\mu^\text{eff}_{\perp} }{\epsilon_0}&=&    \mu_m \frac{1-\left(1- \alpha ^2\right) v^2c_m^{-2}c_0^{-2}}{1-v^2c_m^{-2}c_0^{-2}}
  \\ 
  \label{eq:effxistrat}
\xi^\text{eff}c_0 &=&  \alpha ^2   \frac{v c_0^{-1}c_m^{-2}}{ 1-v^2c_m^{-2}c_0^{-2}}.
\end{eqnarray}
Here $c_m=1/\sqrt{\epsilon_m\mu_m}$ is the relative wave velocity in the unmodulated background medium. From the effective parameters we can obtain the effective wave velocities as, 
\begin{align}
\label{eq:effvelstrat}
    v_\text{eff}^{\pm} = \pm c_0 c_m\frac{1\mp vc_m^{-1}c_0^{-1}}{1\mp vc_m^{-1}c_0^{-1}(1-\alpha^2)}.
\end{align}
From the above we see that if the modulation is only spatial ($v=0$), $v_\text{eff}^{\pm}= \pm c_mc_0 $. In fact, this is a particular case of $v_\text{eff}^{\pm}=\pm d/(d_1/c_1+d_2/c_2)$, the conventional homogenisation result for space-only stratified media, with $c_1=1/\sqrt{\epsilon_1\mu_1}$ and  $c_2=1/\sqrt{\epsilon_2\mu_2}$ being the relative wave velocities in each of the layers. On the other hand, if the modulation is only temporal ($v\rightarrow\infty$), $v_\text{eff}^{\pm}\rightarrow \pm c_mc_0/(1-\alpha^2) $, which is a particular case of $v_\text{eff}^{\pm}=\pm (c_1 d_1+d_2 c_2)/d $, the average wave velocity for time-only stratified media \cite{Pacheco2020}. Given that the system is matched, these analytical results give the exact photonic band-structure at any frequency. On the other hand, we note that although the effective parameters can be calculated for any modulation speed (except for the pole in Eqs. \ref{eq:effepsstrat}-\ref{eq:effvelstrat}), the problem of a travelling stratified medium is ill-defined within the speed range limited by the group velocity of waves in each layer, $c_1<v/c_0<c_2$. In fact, when the modulation speed equals the group velocity of waves in any of the crystal layers, $v/c_0=c_{1,2}$, the effective parameters in the co-moving frame diverge, see Eqs. \ref{eq:epsperpcomovstr}-\ref{eq:xicomovstr}, which results in singularities in the effective parameters that, while being removable, mark the limits of a range of velocities where there are exponentially growing solutions and homogenisation is not valid. 

The expression for the effective wave velocities, Eq. \ref{eq:effvelstrat}, reveals that away from the space-only or time-only modulations, forward and backward modes are affected very differently by the space-time modulation. Figure \ref{fig:stratified} shows the relative effective parameters of this system with $\alpha=0.1$. The effective group velocity in the rest frame, $v_\text{eff}$, is shown in panel (a) for the forward (blue, left axis) and backward (orange, right axis) waves, while panel (b) presents the effective permittivity, permeability and magnetoelectric coupling. The range $c_1<v/c_0<c_2$ is marked with a shaded area. At zero modulation speed, forward and backward modes start at $\pm c_m c_0$. As the modulation speed increases, the effective velocity of backward waves changes very little and monotonously up to $-c_m c_0/(1-\alpha^2)$ at $v\rightarrow\infty$. In fact, this is a consequence of the back-reflection free condition: backward waves interact very little with the modulation. On the other hand, forward waves are strongly affected by it. 
In the subluminal regime, $v<c_1c_0$, the effective forward wave velocity is smaller than the wave velocity in the background medium, $ v_\text{eff}^+<c_mc_0$, and it decreases as the modulation speed increases. This is consistent with the effective refractive index and the magnetoelectric coupling increasing with modulation speed, see panel (b). On the other hand, when the singularity is crossed and the modulation is superluminal, $v>c_2c_0$, the effective velocity changes from being smaller to being larger than the background wave velocity, and then it decreases again up to the limiting value as the modulation speed decreases, consistent with effective parameters smaller than the background parameters, and negative magnetoelectric coupling, see panel (b). 
Additionally, we note here that the effective mode velocity for bilayer space-time crystals was derived through a different method in Ref. \onlinecite{Deck2019uniformvelocity}. 

Having considered a matched space-time modulation, we now look at general values of $(\alpha_e,\alpha_m)=\sqrt{2}\alpha(\cos\phi,\sin\phi)$, while keeping all the parameters the same and assuming low frequencies. For $\phi=\pi/4$, we recover the matched case studied above, $\alpha_{e,m}=\alpha$. Specifically, in Fig. \ref{fig:stratified}(c) we show the effective magnetoelectric coupling, $\xi^\text{eff}$, as a function of modulation speed and $\phi$, the parametrization angle that determines the values of $(\alpha_e,\alpha_m)$. Shaded in gray is the range of modulation speeds where homogenisation is not valid, $c_1<v/c_0<c_2$, which is widest when the system is matched ($\alpha_{e,m}=\pm\alpha $), and vanishes to a point when the electric and magnetic modulations are completely out of phase with each other, ($\pm\alpha_{e}=\mp\alpha_m $). The behaviour of $\xi^\text{eff}$ clearly shows that, if only one of the parameters is modulated, that is, if either $\alpha_e$ or $\alpha_m$ are 0 ($\phi=n\pi/2$, $n=0,1,\cdots$), the system is reciprocal since $\xi^\text{eff}=0$. Conversely, when $\alpha_e=\pm\alpha_m$ ($\phi=n\pi/4$, $n=1,\cdots$), non-reciprocity is maximum as $\xi^\text{eff}$ is maximum. As $\phi$ changes and the relative sign between the electrical and magnetic modulations change, $\xi^\text{eff}$ changes sign. Furthermore, the sign of the effective  magnetoelectric coupling changes when the modulation speed goes from sub- ($v<c_1$) to super-luminal ($v>c_2$) for any phase between the electric and magnetic modulations. It is interesting to note that when the electric and magnetic modulations are completely out of phase and the region where homogenisation is not valid shrinks to a point, the magnetoelectric coupling increases considerably, giving rise to large non-reciprocal effects. 
The results discussed in this section prove that non-reciprocity can be tuned in space-time modulated stratified media by changing the  modulation speed, or the phase between electric and magnetic modulations.

\section{Sinusoidal travelling-wave modulations}
We now consider a sinusoidal travelling-wave modulation,
\begin{eqnarray}
    \epsilon(x,t) &=& \epsilon_m\epsilon_0[1 + 2\alpha_e\cos(gx-\Omega t)] ,\\
    \mu(x,t)      &=& \mu_m\mu_0[1 + 2\alpha_m\cos(gx-\Omega t)],
\end{eqnarray}
where $g$ and $\Omega$ are the spatial and temporal frequencies, $\alpha_{e,m}$ are the electric and magnetic modulation strengths, and $\epsilon_m$ and $\mu_m$ are the background relative permittivity and permeability of the medium. The profile moves with a phase velocity of $v=\Omega/g$. In a previous work, we argued that these metamaterials mimic the relativistic Fresnel drag of light without the need for any material motion \cite{Huidobro2019}. Through a perturbative approach, we derived effective bianisotropic parameters, accurate for small modulation strengths and low modulation speeds. Here we employ the framework developed in this work to derive the exact metamaterial parameters and give an exact formula for the Fresnel drag of light in space-time modulated metamaterials.  

\begin{figure}
    \centering
    \includegraphics{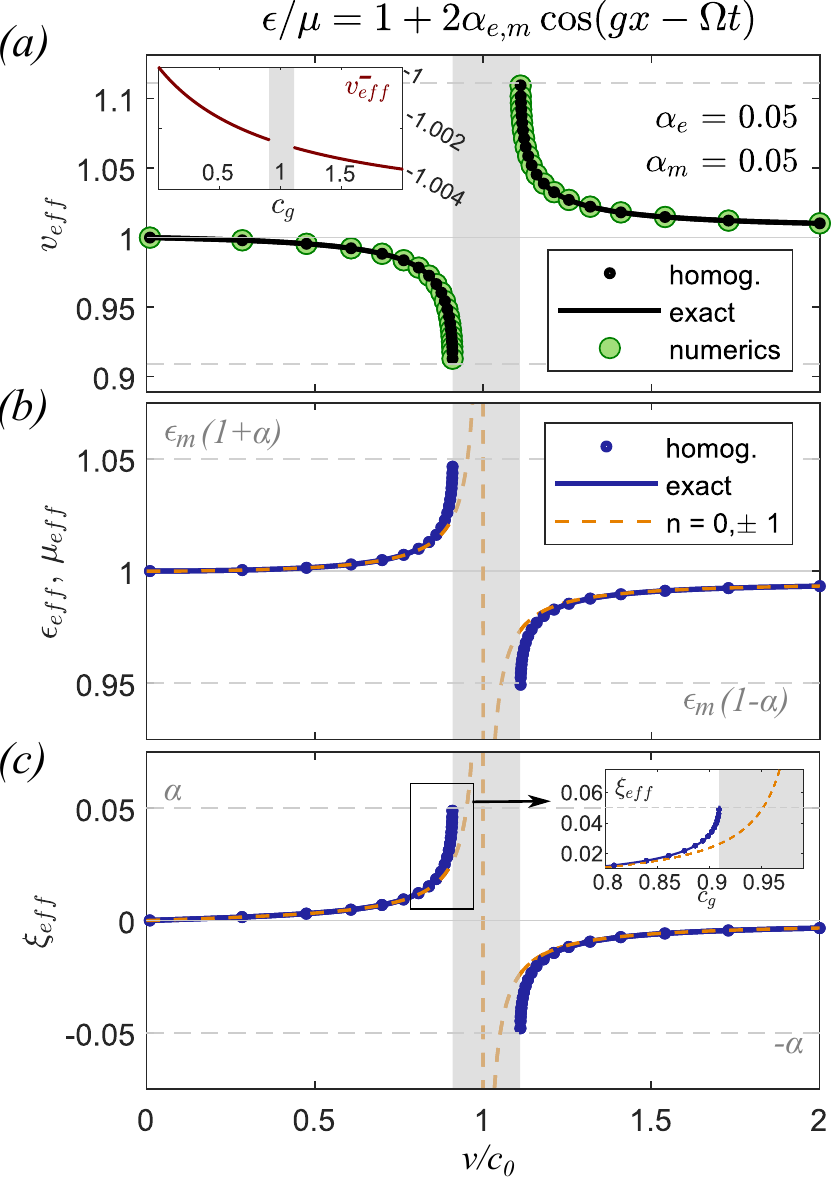}
    \caption{Group velocity and effective parameters for impedance-matched sinusoidal space-time modulations ($\alpha_e=\alpha_m=0.05$) as a function of modulation speed, $v$. The shaded area represents the unstable regime where a band description looses meaning. (a) group velocity, $v_\text{eff}$ from exact analytical theory (black line), homogenisation (black dots), and Floquet-Bloch mode expansion numerics (green circles). (b,c) Effective permittivity and permeability, $\epsilon_\perp^\text{eff}=\mu_\perp^\text{eff}$ (b) and magnetoelectric coupling $\xi_\text{eff}$ (c). Numerical homogenisation (blue dots) and exact formula (blue line) are compared against a three-mode approximation in the Floquet-Bloch expansion (dashed orange line). In all panels, the limiting values of the effective parameters at the threshold of the luminal regime are depicted with gray dashed horizontal lines. We take $\epsilon_m=\mu_m=1$. We show relative values of the wave velocity and effective parameters, i.e., $v_\text{eff}$ is in units of $c_0$, $\epsilon_\text{eff} $ and $\mu_\text{eff} $ in units of $\epsilon_0 $ and $\mu_0$, respectively, and $\xi_\text{eff} $ in units of $c_0^{-1}$. }
    \label{fig:sinusoidalmodulations}
\end{figure}{}

We first consider the case of impedance-matched space-time modulations ($\alpha_{e,m}=\alpha$), where the theory is exact, and we make use of the expression for the effective wave speed in the co-moving frame. From Eq. \ref{eq:effkpm}, we have, for waves co-propagating with the modulation,
\begin{align}
   \frac{1}{\overline{v}'_{+}} &=  \frac{g}{2\pi}\int_0^{2\pi/g} \frac{1}{c_mc_0 [1+2\alpha\cos(gx')]^{-1} - v} \text{d}x' \nonumber \\
   \label{eq:effvelsinusoidalmod} 
   &= -\frac{1}{v} \pm \frac{1}{v\sqrt{(1-4\alpha^2)(v-v_c^+)(v-v_c^-)} }
\end{align}
where we have introduced $v_c^\pm=c_m c_0(1\pm2\alpha)^{-1}$. These modulation speed values correspond to the two critical points where $1/\sqrt{v-v_c^{\pm}} \rightarrow \infty $, and hence the effective wave velocity in the co-moving frame goes to zero. Transforming to the rest frame, $v^+_\text{eff} = \overline{v}'_{+} +v $, we see that at the critical points the effective wave velocity equals the modulation speed, $v^+_\text{eff} = v_c $. In the above expression, $+$ ($-$) corresponds to subluminal (superluminal) modulations, which are bounded by the critical modulation speed values: $v < c_m c_0(1+2\alpha)^{-1} $ corresponds to the subluminal regime, and $ v > c_m c_0(1-2\alpha)^{-1}$ to the superluminal one. In fact, these critical points distinguish the onset of the luminal regime, $c_m c_0(1+2\alpha)^{-1} < v < c_m c_0(1-2\alpha)^{-1}$, where the solution to the integral is imaginary, implying from Eq. \ref{eq:analyticalfieldPDE} that the fields can increase without bound. This is in agreement with the previously identified instability region based on Floquet-Bloch theory for modulations of $\epsilon$ only \cite{cassedy1963dispersion}, or $\epsilon$ and $\mu$ \cite{Taravati2018}, where unidirectional amplification is possible \cite{galiffi2019}. 


In Fig. \ref{fig:sinusoidalmodulations}, we present results for travelling wave impedance-matched space-time modulations with $\alpha=0.05$ and $\epsilon_m=\mu_m=1$. Panel (a) shows the rest frame effective wave velocity, $v_\text{eff}/c_0$, as a function of modulation speed, $v/c_0$, from the exact expression with a solid line. We also show numerical results from Floquet-Bloch theory with green dots, to confirm the accuracy of the result (see Ref. \cite{Huidobro2019} for details on this approach), as well as numerical evaluation of the homogenisation integrals with black dots. Starting at zero modulation speed, where $v_\text{eff}^+=c_0 $ since $c_m=1$, the forward-wave effective velocity decreases down to a threshold value $v_\text{eff}^+=c_m c_0(1+2\alpha)^{-1}\approx 0.91c_0$ when the modulation speed reaches the subluminal critical point, $v=c_m c_0(1+2\alpha)^{-1}$. The decrease in effective wave velocity is accompanied by an increase in effective permittivity and permeability, as well as magnetoelectric coupling. On the other hand, after the luminal region, at the superluminal critical point, $v=c_m c_0(1-2\alpha)^{-1}$, the forward-wave effective velocity takes a limiting value $v_\text{eff}^+=c_m c_0(1-2\alpha)^{-1}\approx 1.11c_0$, and then decreases as the modulation velocity increases, approaching $v_\text{eff}^+=c_0(1+2\alpha^2) $. Differently from the travelling stratified crystal, in this case there is a saturation in the value of wave group velocity at the lower and upper threshold of the range where homogenisation is not valid.

Figure \ref{fig:sinusoidalmodulations} (b,c) shows the effective permittivity and permeability ($\epsilon_\perp^\text{eff}/\epsilon_0 = \mu_\perp^\text{eff}/\mu_0 $, b) and effective magnetoelectric coupling, ($\xi^\text{eff} c_0$, c). The homogenisation integrals, which we showed are exact for impedance-matched systems, can be solved analytically and we use them to plot the effective parameters with a blue line, comparing also to numerically evaluated integrals (plotted with dots). For the matched case under consideration, our method yields, 
\begin{eqnarray}
    \epsilon_\perp^\text{eff} &=& \epsilon_m \epsilon_0 \frac{1}{2vc_0^{-1}c_m^{-1}} \frac{\Gamma_- \mp \Gamma_+}{\Gamma_-\Gamma_+} \\ 
    \mu_\perp^\text{eff} &=& \mu_m \mu_0 \frac{1}{2vc_0^{-1}c_m^{-1}} \frac{\Gamma_- \mp \Gamma_+}{\Gamma_-\Gamma_+} \\ 
     \xi^\text{eff} c_0 &=& -\frac{1}{2vc_0^{-1}} \frac{\Gamma_- \mp \Gamma_+ - 2 \Gamma_-\Gamma_+}{\Gamma_-\Gamma_+} \\ 
\end{eqnarray}
where the top (bottom) sign corresponds to subluminal (superluminal) speeds and we have introduced the shorthand 
\begin{equation}
    \Gamma_{\pm} = \frac{c_mc_0}{\sqrt{(1-4\alpha^2) (v\pm v_c^+)(v\pm v_c^-)}}.
\end{equation} 
We note that the homogenisation integrals can be solved analytically for general $\alpha_e$ and $\alpha_m$, not only for the matched case, and we give the general expressions in the S.M. In agreement with the effective wave velocity, the effective permittivity and permeability increase above the background value, and the magnetoelectric coupling increases above zero, as the modulation speed increases from 0 to the subluminal critical speed, $v=c_mc_0(1+2\alpha)^{-1}$, where $\epsilon_\perp^\text{eff}/\epsilon_m\epsilon_0 = \mu_\perp^\text{eff}/\mu_m\mu_0 = 1+\alpha $, and $\xi^\text{eff} c_0 = \alpha$. Then, after the luminal region, the permittivity and permeability are reduced below their background values, and the magnetoelectric coupling changes sign. They start at threshold values $\epsilon_\perp^\text{eff}/\epsilon_m\epsilon_0 = \mu_\perp^\text{eff}/\mu_m\mu_0 = 1-\alpha $, and $\xi^\text{eff} c_0 = -\alpha$ when the modulation speed equals the superluminal critical velocity, $v=c_mc_0(1+2\alpha)^{-1}$. As the modulation speed increases, they increase approaching the limiting values at infinite modulation speed, $\epsilon_\perp^\text{eff}/\epsilon_0\rightarrow\epsilon_m(1-2\alpha^2)$, $\mu_\perp^\text{eff}/\mu_0\rightarrow\mu_m(1-2\alpha^2)$ and $\xi^\text{eff}\rightarrow0$. In addition, panels (b,c) also show results for the effective parameters (dashed orange line) obtained from a perturbative approach that includes three-modes in a Floquet-Bloch expansion. As detailed in Ref. \cite{Huidobro2019}, this is a good approximation for small modulation strengths, $\alpha\ll 1$, and modulation speeds far from the luminal region, $v\ll c_mc_0$ or $v\gg c_mc_0$. Indeed, we can see in the plot how the perturbative result is very accurate for these low and high velocities. However, it completely fails to predict the correct behaviour close to the luminal regime, and in particular, it misses the saturation of the effective parameters at the critical modulation speeds. These critical points represent a transition between a system described accurately in a Bloch wave picture, and the amplification regime, and will be studied elsewhere.

\begin{figure}
    \centering
    \includegraphics{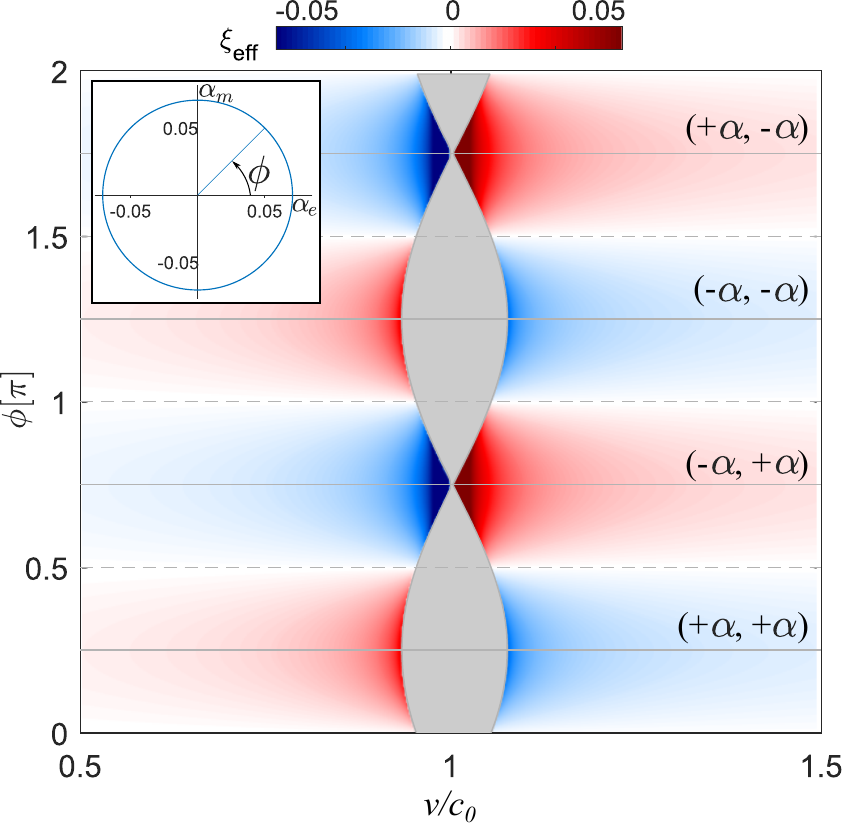}
    \caption{Effective magnetoelectric coupling as a function of space-time modulation speed, $v/c_0$, and the values of $(\alpha_e,\alpha_m)$, parametrized by $\phi$ (see inset). Effective $\xi$ is zero whenever only one of the parameters is modulated (marked with horizontal dashed lines). The grey region centered at $v=c_0$ corresponds to the luminal regime, which is maximised when the system is matched $(\alpha_e,\alpha_m) = (\pm\alpha,\pm\alpha)$, and minimised when the modulations oppose each other $(\alpha_e,\alpha_m) = (\pm\alpha,\mp\alpha)$ (all four cases marked with solid horizontal lines). We take $\alpha=0.05$, $\epsilon_m=\mu_m=1$, and $c_m=1$, as in Fig. 3. We use units of $c_0^{-1}$ for $\xi_\text{eff}$. }
    \label{fig:magnetoelectricmap}
\end{figure}{}

Let us now consider general space-time modulations of the permittivity and permeability. We parametrize the electric and magnetic modulations through $(\alpha_e,\alpha_m)=\sqrt{2}\alpha(\cos\phi,\sin\phi)$, and we take $\alpha=0.05$, such that when $\phi=\pi/4$, we have $\alpha_{e,m}=\alpha$ as in the previously studied case. Figure \ref{fig:magnetoelectricmap} shows a contour plot of the effective magnetoelectric coupling as a function of the modulation speed, $v/c_0$, and the parametrization angle, $\phi$, similar to what we showed for the travelling stratified medium in Fig. \ref{fig:stratified}. 
The luminal regime is given by the range of modulation speeds limited by the minimum and maximum loacl group velocities, 
\begin{align}
    \frac{c_m}{\sqrt{(1+2\alpha_e)(1+2\alpha_m)}} \leq \frac{v}{c_0} \leq \frac{c_m}{\sqrt{(1-2\alpha_e)(1-2\alpha_m)}}.
\end{align}
This range, which corresponds to the shaded gray area in Fig. 4, was previously identified for the case of $\epsilon$ and $\mu$ modulations from Floquet-Bloch theory as the region where a band description of the system fails \cite{Taravati2018}, while here it stems from our analytical treatment. Intuitively, the luminal regime can be understood as the range of modulation velocities bounded by the minimum and maximum local phase velocities of the modulation in the co-moving frame $\min [c(x')] \leq v \leq \max [c(x')]$. As can be seen in Fig. \ref{fig:magnetoelectricmap}, this criterion implies that the luminal regime is widest for the matched case studied above, $\alpha_e=\alpha_m=\pm\alpha$, while it is minimum (width of order $\alpha^2$) when the modulations are of the same size but out of phase $\alpha_e=-\alpha_m=\pm\alpha$, and its size varies between these extreme cases. While an analytical treatment of this regime is also possible \cite{pendry2020new}, here we concentrate on parameters outside of the luminal regime, where the system can be represented by effective material parameters, which as we showed are exact in the low frequency limit. Looking at the value of $\xi_\text{eff}$, it can be seen that it is non-zero only when both $\alpha_e$ and $\alpha_m$ are non-zero. This implies that whenever both the permittivity and permeability are modulated, the system is non-reciprocal at zero frequency \cite{Huidobro2019}. In addition, we see how the effective magnetoelectric coupling is largest in size at the lower and upper thresholds of the luminal regime, and for the phases where the luminal range is widest, that is, for electrical and magnetic modulations of the same size, matched, $(\pm\alpha,\pm\alpha)$ or in anti-phase,  $(\pm\alpha,\mp\alpha)$. Its sign changes between the subluminal and superluminal regime, and also when the relative sign between $\alpha_e$ and $\alpha_m$ changes, revealing the possibility of tuning the nonreciprocity direction by tuning the modulation speed, or the phase of the electric and magnetic modulations. To complement the study of the magnetoelectric coupling, in the S.M. we present results for all effective parameters for an instance of non-matched modulations.




Finally, we make a connection to the Fresnel drag of light, by establishing an exact mapping between the bianisotropic metamaterial and an equivalent (non-bianisotropic) moving medium. The non-reciprocal dispersion curves of space-time modulations of both the electric and magnetic parameters can be linked to the relativistic dragging of light by moving matter, even though there is no physical motion \cite{Huidobro2019}. In particular, the effective bianisotropic medium characterised by the parameters given in Eqs. (24-28) can be mapped to an uniaxial medium with permittivity and permeability tensors, 
\begin{eqnarray}
    \boldsymbol{\epsilon}_\text{eq} &=& \left[ 
    \begin{array}{ccc} 
        \epsilon_{\text{eq,}||} & 0 & 0 \\
        0 & \epsilon_{\text{eq,}\perp} & 0 \\
        0 & 0 & \epsilon_{\text{eq,}\perp}
    \end{array} \right], \\
    \boldsymbol{\mu}_\text{eq} &=& \left[ 
    \begin{array}{ccc} 
        \mu_{\text{eq,}||} & 0 & 0 \\
        0 & \mu_{\text{eq,}\perp} & 0 \\
        0 & 0 & \mu_{\text{eq,}\perp}
    \end{array} \right],
\end{eqnarray} 
moving with velocity $v_D$. From a Lorentz transformation between both frames we have, 
\begin{eqnarray} 
    \epsilon_\perp^\text{eff} &=& \epsilon_{\text{eq,}\perp}\frac{1-v_D^2/c_0^2}{1-\epsilon_{\text{eq,}\perp}^2\mu_\perp^\text{eff} v_D^2/\epsilon_\perp^\text{eff}} \\
    \xi^\text{eff} &=& \frac{v_D}{c_0^2}\frac{\epsilon_\perp'^2\mu_\perp^\text{eff}/\epsilon_\perp^\text{eff}c_0^2 -1}{1-\epsilon_{\text{eq,}\perp}^2\mu_\perp^\text{eff} v_D^2/\epsilon_\perp^\text{eff}},
\end{eqnarray}
together with $\mu_{\text{eq,}\perp} = \epsilon_{\text{eq,}\perp} \mu_\perp^\text{eff} / \epsilon_\perp^\text{eff}$ and  $\epsilon_{\text{eq,}||} =\epsilon_{||}^\text{eff} $, $\mu_{\text{eq,}||} =\mu_{||}^\text{eff} $.  We can then solve the above system for $v_D$, $\epsilon_{\text{eq,}\perp}$ and $\mu_{\text{eq,}\perp}$, thus completely characterising the equivalent moving medium. The obtained analytical expressions are given in the S.M..

\begin{figure}[t]
    \centering
    \includegraphics{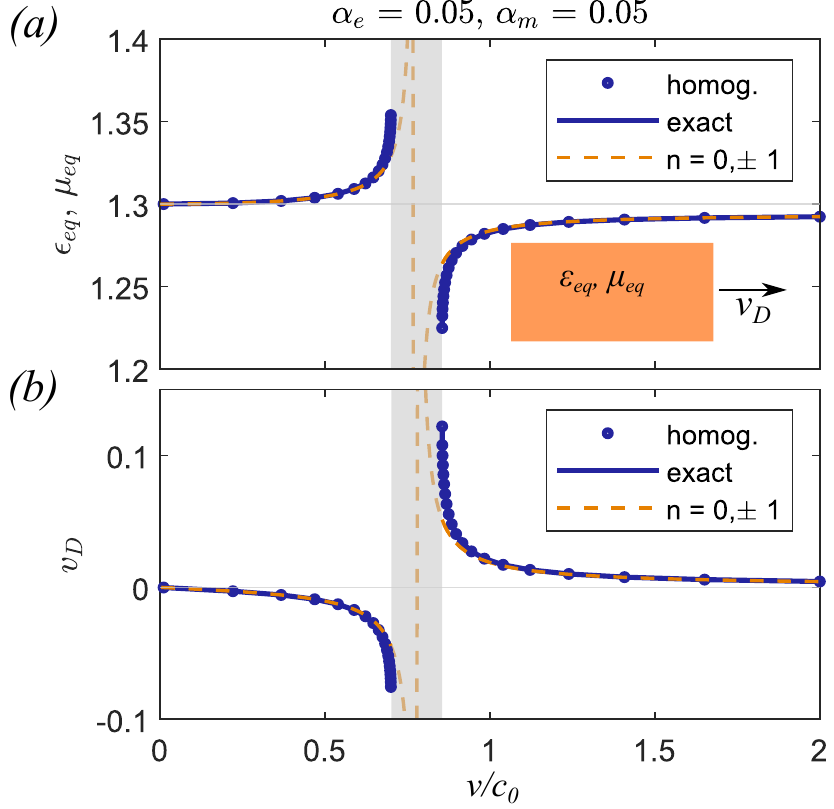}
    \caption{Electromagnetic parameters, permittivity and permeability (a), and  velocity, $v_D$ (b), of the equivalent uniaxial moving medium as a function of speed modulation. Numerical homogenisation (blue dots) and exact formula (blue line) are compared against a three-mode approximation in the Floquet-Bloch expansion (dashed orange line). We take $\epsilon_m=\mu_m=1.3$ and $\alpha_e=\alpha_m=0.05$. The permittivity and permeability are given in units of $\epsilon_0$ and $\mu_0$, respectively, and the drag velocity in units of $c_0$. }
    \label{fig:Fresneldrag}
\end{figure}{}

Figure \ref{fig:Fresneldrag} shows the equivalent moving medium parameters and velocity for a space-time modulated system with parameters $\epsilon_m=\mu_m=1.3 $, $\alpha_e=\alpha_m=0.05$. For this impedance matched case, we can exploit the exact effective parameters to derive an exact mapping to an equivalent moving medium, shown as blue lines. We also plot with dashed orange lines results obtained from a Floquet-Bloch expansion assuming three-modes for comparison \cite{Huidobro2019}. Similar to the behaviour of the effective parameters, the equivalent moving medium permittivity and permeability (a) and velocity (b), are only defined outside of the luminal range (shaded area). Starting at zero modulation velocity, the equivalent medium is isotropic and stationary, and as the modulation speed increases, the equivalent medium effective parameters increase above the background values, up to a threshold value at the critical subluminal speed that the perturbation theory approach fails to capture (dashed orange line). The equivalent medium velocity is negative, and increases in size also down to a threshold value. On the other hand, after crossing the luminal range, the equivalent medium parameters flip to a threshold value below the background parameters at the superluminal critical speed, and then approach the background parameters as the modulation speed increases. The velocity of the equivalent moving medium turns positive, and decreases towards zero as the modulation approaches a temporal-only modulation. Thus, the drag direction can be switched by switching between sub- and superluminal modulation speeds, while keeping the same modulation direction. The different sign of the drag velocity is linked to the opposite signs of the effective magnetoelectric coupling for sub- and superluminal modulations. In brief, subluminal (superluminal) gratings, slow down (speed up) forward waves travelling through the modulated medium, and result in equifrequency contours displaced in the same (opposite) direction as the modulation phase velocity \cite{Huidobro2019}. These exact results prove the link between the Fresnel drag of light in moving matter and space-time modulated media, where there is no physical motion.

Finally, it is worth stressing the different equivalences that we have shown in this paper: \textit{(i)} Initially we consider an isotropic medium subject to a spatio-temporal modulation of its permittivity and permeability, \textit{(ii)} Through a Galilean transformation, the space-time metamaterial maps to a bianisotropic medium in a frame co-moving with the modulation at speed $v$ (Eqs. 21-23), \textit{(iii)} In the rest frame the spatio-temporal metamaterial maps to a bianisotropic effective medium (Eqs. 24-28), and \textit{(iv)} A Lorentz frame moving at velocity $v_D$ with respect to the laboratory frame can be found where the medium is uniaxial (non-bianisotropic), giving rise to the explanation of the emerging bianisotropy as a Fresnel drag effect.


\section{Conclusions}
To summarize, here we have presented the first homogenisation theory of space-time metamaterials, and we have shown it is exact in the absence of back-scattering, that is, in the case of impedance-matched modulations, or at low frequencies when this condition is lifted. Our theory provides analytical expressions for the effective medium parameters of travelling space-time modulations. While we have considered the spatio-temporal modulation of an isotropic medium, the theory can be extended to more complex scenarios such as the space-time modulation of non-isotropic media, and shed light on topological transitions or exceptional points that may emerge there. In addition, having focused here on Maxwell's equations, our framework can be extended to other wave theories. 

We have looked in detail at two instances of travelling space-time media: a stratified crystal, and a sinusoidal grating. These constitute the two extreme cases of travelling modulations of one Fourier component (sine wave), and a square wave, and while they yield similar phenomenology at low and high modulation speeds, there are differences in their behaviour at sub- and super-luminal velocities that are close to the velocity of light. While for the stratified crystal the effective medium parameters can in principle be calculated within the luminal range where there are no stable solutions, in the case of sinusoidal modulations the effective medium parameters saturate at the edges of the singular regime. 
We expect this critical behaviour to lead to rich physics in space-time modulated metamaterials. Furthermore, our analysis proves that  space-time media based on travelling wave modulations are exactly equivalent (outside the luminal range) to a moving uniaxial uniform material in the long wavelength regime, with an equivalent velocity of motion that is not the same as the metamaterial modulation speed.

\begin{acknowledgments}
P.A.H. and M.S. acknowledge funding from Funda\c c\~ao para a Ci\^encia e a Tecnologia and Instituto de Telecomunica\c c\~oes under project UID/50008/2020. P.A.H. is supported by the CEEC Individual program from Funda\c c\~ao para a Ci\^encia e a Tecnologia with reference CEECIND/03866/2017. E.G. acknowledges support through a studentship in the Centre for Doctoral Training on Theory and Simulation of Materials at Imperial College London funded by the EPSRC (EP/L015579/1). J.B.P. acknowledges funding from the Gordon and Betty Moore Foundation.
\end{acknowledgments}

\appendix

\bibliography{main}



\end{document}


\title{Supplemental Material:\\ Homogenisation Theory of Space-Time Metamaterials}

\author{P. A. Huidobro}
 \email{p.arroyo-huidobro@lx.it.pt}
\affiliation{Instituto de Telecomunica\c c\~oes, Instituto Superior Tecnico-University of Lisbon, Avenida Rovisco Pais 1, Lisboa, 1049‐001 Portugal
}
\author{M.G. Silveirinha}%
\affiliation{Instituto de Telecomunica\c c\~oes, Instituto Superior Tecnico-University of Lisbon, Avenida Rovisco Pais 1, Lisboa, 1049‐001 Portugal
}%

\author{E. Galiffi}
\affiliation{The Blackett Laboratory, Department of Physics, Imperial College London, London, SW7 2AZ UK}%
\author{J.B. Pendry}
\affiliation{The Blackett Laboratory, Department of Physics, Imperial College London, London, SW7 2AZ UK}%

\maketitle

\tableofcontents

\section{Homogenisation Theory}
\subsection{Derivation of the constitutive relations in the co-moving frame}
\label{sec:homogcomoving}
Here we use Lorentz's transformations to write the constitutive relations of the electromagnetic fields in a frame co-moving with the modulations. At the end we particularize to a Galilean transformation.  

In the lab frame, we have Maxwell's equations, 
\begin{eqnarray}
    \nabla\times\mathbf{E} &=& -\frac{\partial \mathbf{B}}{\partial t}, \\
    \nabla\times\mathbf{H} &=& \frac{\partial \mathbf{D}}{\partial t},
\end{eqnarray}
with constitutive relations, 
\begin{eqnarray}
    \label{eq:conseqslab1}
    \mathbf{D} &=& \epsilon(x-v t) \mathbf{E},\\
    \label{eq:conseqslab2}
    \mathbf{B} &=& \mu(x-v t)  \mathbf{H}.
\end{eqnarray}
Here, $\epsilon$, $\mu$, include $\epsilon_0$,$\mu_0$, such that $\epsilon\mu$ has units of $1/c^2$. 
Maxwell's equations are form-invariant under Lorentz transformations,
\begin{eqnarray}
    x'&=& \gamma(x-v t), \\
    t'&=& \gamma(t-\frac{v}{c^2} x),
\end{eqnarray}
with $\gamma=(1-v^2/c^2)^{1/2}$
so in a (primed) frame co-moving with the modulations at speed $\mathbf{v}=v\mathbf{x}$, 
\begin{eqnarray}
    \nabla'\times\mathbf{E'} &=& -\frac{\partial \mathbf{B'}}{\partial t'}, \\
    \nabla'\times\mathbf{H'} &=& \frac{\partial \mathbf{D'}}{\partial t'}.
\end{eqnarray}

In order to write the constitutive relations in the co-moving frame we start from Lorentz's transformations of the electromagnetic fields \cite{Kong}, 
\begin{eqnarray}
    \mathbf{E}_{||}' &=& \mathbf{E}_{||}, \\
    \mathbf{B}_{||}' &=& \mathbf{B}_{||}, \\
    \mathbf{E}_{\perp}' &=& \gamma\left( \mathbf{E}_{\perp} + \mathbf{v}\times \mathbf{B}_{\perp} \right), \\
    \mathbf{B}_{\perp}' &=& \gamma\left( \mathbf{B}_{\perp} -\frac{1}{c^2} \mathbf{v}\times \mathbf{E}_{\perp} \right),
\end{eqnarray}
and
\begin{eqnarray}
    \mathbf{D}_{||}' &=& \mathbf{D}_{||}, \\
    \mathbf{H}_{||}' &=& \mathbf{H}_{||}, \\
    \mathbf{D}_{\perp}' &=& \gamma\left( \mathbf{D}_{\perp} + \frac{1}{c^2} \mathbf{v}\times \mathbf{H}_{\perp} \right), \\
    \mathbf{H}_{\perp}' &=& \gamma\left( \mathbf{H}_{\perp} -\mathbf{v}\times \mathbf{D}_{\perp} \right).
\end{eqnarray}

Thus, and from Eqs. \ref{eq:conseqslab1}-\ref{eq:conseqslab2}, we have in the co-moving frame, 
\begin{eqnarray}
\label{eq:conseqscomovpar}
    \mathbf{D}_{||}' &=& \epsilon(x'/\gamma)  \mathbf{E}_{||}',\\
    \mathbf{B}_{||}' &=& \mu(x'/\gamma)  \mathbf{H}_{||}'.
\end{eqnarray}
On the other hand, for the components in the plane perpendicular to the velocity, we can write 
\begin{eqnarray}
    \begin{bmatrix}
         \mathbf{D}_{\perp}' \\ \mathbf{B}_{\perp}'
    \end{bmatrix}  = \gamma
    \begin{bmatrix}
         \epsilon & \frac{1}{c^2}\mathbf{v}\times\mathbbm{1}\\ -\frac{1}{c^2}\mathbf{v}\times\mathbbm{1} & \mu
    \end{bmatrix}
    \begin{bmatrix}
         \mathbf{E}_{\perp} \\ \mathbf{H}_{\perp}
    \end{bmatrix}
\end{eqnarray}
\begin{eqnarray}
    \begin{bmatrix}
         \mathbf{E}_{\perp} \\ \mathbf{H}_{\perp}
    \end{bmatrix}  = \gamma  \begin{bmatrix}
         \mathbf{E}_{\perp}' \\ \mathbf{H}_{\perp}'
    \end{bmatrix} +\gamma
    \begin{bmatrix}
         0& -\mathbf{v}\times\mathbbm{1}\\ \mathbf{v}\times\mathbbm{1} & 0
    \end{bmatrix}
    \begin{bmatrix}
         \mathbf{D}_{\perp}' \\ \mathbf{B}_{\perp}'
    \end{bmatrix}
\end{eqnarray}
which after some manipulations yields, 
\begin{eqnarray}
    \begin{bmatrix}{}
        \mathbf{D'}_{\perp} \\
         \mathbf{B'}_{\perp}
    \end{bmatrix} &=& \frac{1}{1-\epsilon(x'/\gamma)\mu(x'/\gamma) v^2} \times
    \nonumber \\
    &&\begin{bmatrix}
        \gamma^{-2} \epsilon(x'/\gamma) \mathbbm{1}& -\left[\epsilon(x')\mu(x')c^2 - 1\right] \frac{\mathbf{v}}{c^2}\times\mathbbm{1}\\ 
        \left[\epsilon(x')\mu(x')c^2 - 1\right] \frac{\mathbf{v}}{c^2}\times\mathbbm{1} & \gamma^{-2} \mu(x'/\gamma) \mathbbm{1} 
    \end{bmatrix}
    \begin{bmatrix}{}
        \mathbf{E'}_{\perp} \\
         \mathbf{H'}_{\perp}
    \end{bmatrix} 
\end{eqnarray}{}

It is interesting to note that in the previous formulas $c$ may be understood as a free-parameter, in the sense that the considered field mapping transforms the original Maxwell's equations for the unprimed fields into an equivalent set of equations for the primed fields with the constitutive relations (S21). The choice $c=1/\sqrt{\epsilon_0\mu_0}$ corresponds to the usual Lorentz transformation and ensures that the vacuum constitutive relations are the same for the primed and unprimed fields. For our purposes, it is however more convenient to pick $c\rightarrow\infty$ ($\gamma=1$), such that the coordinate transformation reduces to Galilean one, 
\begin{eqnarray}
    x'&=&x-v t, \\
    t'&=& t.
\end{eqnarray}
This is more suited for our purposes since we are concerned with modulated media, and not with moving matter, such that the velocity of the modulation is not restricted by the speed of light. With this, we arrive at,
\begin{eqnarray}
\label{eq:conseqscomovparGal}
    \mathbf{D}_{||}' &=& \epsilon(x') \mathbf{E}_{||}',\\
    \mathbf{B}_{||}' &=& \mu(x')  \mathbf{H}_{||}', \\
\label{eq:conseqscomovperpGal}
    \begin{bmatrix}{}
        \mathbf{D'}_{\perp} \\
         \mathbf{B'}_{\perp}
    \end{bmatrix} &=& \frac{1}{1-\epsilon(x')\mu(x') v^2} \begin{bmatrix}
        \epsilon(x') \mathbbm{1}& -\epsilon(x')\mu(x')\mathbf{v}\times\mathbbm{1}\\ 
        \epsilon(x')\mu(x')\mathbf{v}\times\mathbbm{1} & \mu(x') \mathbbm{1} 
    \end{bmatrix}
    \begin{bmatrix}{}
        \mathbf{E'}_{\perp} \\
         \mathbf{H'}_{\perp}
    \end{bmatrix}
\end{eqnarray}{}
These are Eqs. (5,6) of the main text. Equivalently, 
\begin{eqnarray}
    \begin{bmatrix}{}
        \mathbf{D'}_{\perp} \\
         \mathbf{B'}_{\perp}
    \end{bmatrix} &=& \frac{1}{1-v^2/c(x')^2} \begin{bmatrix}
        \epsilon(x') \mathbbm{1}& -c(x')^{-2}\mathbf{v}\times\mathbbm{1}\\ 
        c(x')^{-2}\mathbf{v}\times\mathbbm{1} & \mu(x') \mathbbm{1} 
    \end{bmatrix}
    \begin{bmatrix}{}
        \mathbf{E'}_{\perp} \\
         \mathbf{H'}_{\perp}
    \end{bmatrix} 
\end{eqnarray}{}
with $c(x')=1/\sqrt{\epsilon(x')\mu(x')}$ being the local phase velocity in the co-moving frame. 

The above constitutive equations, Eqs. \ref{eq:conseqscomovparGal}-\ref{eq:conseqscomovperpGal}, depend only on the transformed spatial coordinate, $x'$, and represent bianisotropic medium parameters,  
\begin{eqnarray}
    \label{eq:bianisotropiccomov1}
    \mathbf{D}_{||}' &=& \epsilon_{||}'(x') \mathbf{E}_{||}',\\
    \label{eq:bianisotropiccomov2}
    \mathbf{B}_{||}' &=& \mu_{||}'(x')\mathbf{H}_{||}', \\ 
    \label{eq:bianisotropiccomov3}
    \begin{bmatrix}{}
        \mathbf{D'}_{\perp} \\
         \mathbf{B'}_{\perp}
    \end{bmatrix} &=&   \begin{bmatrix}
       \hat{ \boldsymbol{\epsilon}}_\perp'(x') \mathbbm{1}& \hat{\boldsymbol{\xi'}}_\perp(x')\\ 
       -\hat{\boldsymbol{\xi'}}_\perp(x') & \hat{\boldsymbol{\mu}}_\perp'(x') \mathbbm{1} 
    \end{bmatrix}
    \begin{bmatrix}{}
        \mathbf{E'}_{\perp} \\
         \mathbf{H'}_{\perp}
    \end{bmatrix} 
\end{eqnarray}{} 
with $\mathbf{u}_v$ being the unit vector in the direction of the modulation, and $\hat{\boldsymbol{\xi'}}_\perp = \xi' \mathbf{u}_v\times\mathbbm{1} $. Writing the perpendicular components explicitly, 
\begin{eqnarray}
    \begin{bmatrix}{}
        \mathbf{D'}_{y} \\ \mathbf{D'}_{z} \\
         \mathbf{B'}_{y} \\ \mathbf{B'}_{z}
    \end{bmatrix} &=&   \begin{bmatrix}
        \epsilon_\perp' & 0 & 0 & -\xi' \\ 
        0 &  \epsilon_\perp' & \xi' & 0 \\
       0 & \xi' & \mu_\perp' & 0 \\
      -\xi' & 0 & 0 & \mu_\perp'
    \end{bmatrix}
    \begin{bmatrix}{}
        \mathbf{E'}_{\perp} \\
         \mathbf{H'}_{\perp}
    \end{bmatrix}. 
\end{eqnarray}{} 
The set of effective parameters is obtained by homogenising Eqs. \ref{eq:bianisotropiccomov1}-\ref{eq:bianisotropiccomov3} as detailed in the main text: 
\begin{eqnarray}
    \overline{\epsilon'}_{||} &=& \left[ \frac{1}{d} \int_0^d\frac{1}{\epsilon'_{||}(x')}\,\text{d}x'\right]^{-1} \\ 
    \label{eq:mueff}
     \overline{\mu'}_{||} &=& \left[ \frac{1}{d} \int_0^d\frac{1}{\mu'_{||}(x')}\,\text{d}x'\right]^{-1} ,\\
    \overline{\epsilon'}_{\perp} &=& \frac{1}{d} \int_0^d  \frac{\epsilon(x')}{1-\epsilon(x')\mu(x') v^2} \text{d}x' , \\ 
    \overline{\mu'}_{\perp} &=& \frac{1}{d} \int_0^d  \frac{\mu(x')}{1-\epsilon(x')\mu(x') v^2} \text{d}x' , \\ 
    \overline{\xi'} &=& -v \frac{1}{d} \int_0^d  \frac{\epsilon(x')\mu(x')}{1-\epsilon(x')\mu(x') v^2} \text{d}x' , \\ 
\end{eqnarray}{}
with $d$ the spatial periodicity.

\subsection{Effective parameters in the rest frame}
Once homogenisation has been performed in the co-moving frame (primed parameters), the effective parameters need to be transformed back to the lab frame. Starting from bianisotropic-type equations in the co-moving frame, Eqs. \ref{eq:bianisotropiccomov1}- \ref{eq:bianisotropiccomov3}, in the lab frame we have from Lorentz's transformations for the parallel field components \cite{Kong}, 
\begin{eqnarray}
    \mathbf{D}_{||} &=& \epsilon_{||} \mathbf{E}_{||},\\
    \mathbf{B}_{||} &=& \mu_{||} \mathbf{H}_{||}, \\ 
\end{eqnarray}
On the other hand, for the perpendicular field components, Lorentz's transformations read as,   
\begin{eqnarray}
    \begin{bmatrix}{}
        \mathbf{D}_{\perp} \\
         \mathbf{B}_{\perp}
    \end{bmatrix} &=&  \gamma \begin{bmatrix}{}
        \mathbf{D}_{\perp}' \\
         \mathbf{B}_{\perp}'
    \end{bmatrix} +
    \begin{bmatrix}
        0 & -\frac{v}{c^2}\mathbf{u}_v\times\mathbbm{1} \\ 
       \frac{v}{c^2}\mathbf{u}_v\times\mathbbm{1} & 0 
    \end{bmatrix}
    \begin{bmatrix}{}
        \mathbf{E'}_{\perp} \\
         \mathbf{H'}_{\perp}
    \end{bmatrix} 
\end{eqnarray}{}
\begin{eqnarray}
    \begin{bmatrix}{}
        \mathbf{E'}_{\perp} \\
         \mathbf{H'}_{\perp}
    \end{bmatrix} &=&  \gamma \begin{bmatrix}{}
        \mathbf{E}_{\perp} \\
         \mathbf{H}_{\perp}
    \end{bmatrix} +
    \begin{bmatrix}
        0 & v\mathbf{u}_v\times\mathbbm{1} \\ 
       -v\mathbf{u}_v\times\mathbbm{1} & 0 
    \end{bmatrix}
    \begin{bmatrix}{}
        \mathbf{D}_{\perp} \\
         \mathbf{B}_{\perp}
    \end{bmatrix} 
\end{eqnarray}{}
From the above we can derive the constitutive relations in the lab frame, 
\begin{eqnarray}
   && \begin{bmatrix}{}
        \mathbf{D}_{\perp} \\
         \mathbf{B}_{\perp}
    \end{bmatrix} =  
    \frac{1}{(1-v\xi)^2 - v^2\epsilon'_\perp\mu'_\perp} \times \\\nonumber
    &&\begin{bmatrix}
        \gamma^{-2}\epsilon_\perp\mathbbm{1} & \left[v\epsilon_\perp\mu_\perp + (1-v\xi')\left( \xi' - \frac{v}{c^2}\right) \right]\mathbf{u}_v\times\mathbbm{1} \\ 
       -\left[v\epsilon_\perp\mu_\perp + (1+v\xi')\left( \xi' - \frac{v}{c^2}\right) \right]\mathbf{u}_v\times\mathbbm{1} &  \gamma^{-2}\mu_\perp\mathbbm{1}
    \end{bmatrix}
    \begin{bmatrix}{}
        \mathbf{E}_{\perp} \\
         \mathbf{H}_{\perp}
    \end{bmatrix} 
\end{eqnarray}{}
Taking the Galilean limit, we have 
\begin{eqnarray}
   && \begin{bmatrix}{}
        \mathbf{D}_{\perp} \\
         \mathbf{B}_{\perp}
    \end{bmatrix} =  
    \frac{1}{(1-v\xi)^2 - v^2\epsilon'_\perp\mu'_\perp} \times \\\nonumber
    &&\begin{bmatrix}
        \epsilon_\perp\mathbbm{1} & \left[v\epsilon_\perp\mu_\perp + (1-v\xi')\xi' \right]\mathbf{u}_v\times\mathbbm{1} \\ 
       -\left[v\epsilon_\perp\mu_\perp + (1+v\xi')\xi' \right]\mathbf{u}_v\times\mathbbm{1} &  \mu_\perp\mathbbm{1}
    \end{bmatrix}
    \begin{bmatrix}{}
        \mathbf{E}_{\perp} \\
         \mathbf{H}_{\perp}
    \end{bmatrix} 
\end{eqnarray}{}
where it is clear that the constitutive matrix in the lab frame is also bianisotropic, and we can identify the effective parameters in the lab frame as, 
\begin{eqnarray}
    \mathbf{D}_{||} &=& \epsilon_{||}^\text{eff} \mathbf{E}_{||},\\
    \mathbf{B}_{||} &=& \mu_{||}^\text{eff} \mathbf{H}_{||}, \\ 
    \begin{bmatrix}{}
        \mathbf{D}_{\perp} \\
         \mathbf{B}_{\perp}
    \end{bmatrix} &=& \begin{bmatrix}
        \epsilon_\perp^\text{eff}\mathbbm{1} & \xi^\text{eff}\mathbf{u}_v\times\mathbbm{1} \\ 
       -\xi^\text{eff}\mathbf{u}_v\times\mathbbm{1} &  \mu_\perp^\text{eff}\mathbbm{1}
    \end{bmatrix}
    \begin{bmatrix}{}
        \mathbf{E}_{\perp} \\
         \mathbf{H}_{\perp}
    \end{bmatrix} 
\end{eqnarray}{}

Hence, 
\begin{eqnarray}
    \epsilon_{||}^\text{eff} &=&  \overline{\epsilon'}_{||} , \\
    \mu_{||}^\text{eff} &=&  \overline{\mu'}_{||} , \\
    \epsilon_{\perp}^\text{eff} &=&  \frac{\overline{\epsilon'}_\perp} { (1-v \overline{\xi'})^2 - v^2 \overline{\epsilon'}_\perp \overline{\mu'}_\perp } , \\
    \mu_{\perp}^\text{eff} &=&  \frac{\overline{\mu'}_\perp} { (1-v \overline{\xi'})^2 - v^2 \overline{\epsilon'}_\perp \overline{\mu'}_\perp } , \\ 
    \xi^\text{eff} &=&  \frac{v\overline{\epsilon'}_\perp \overline{\mu'}_\perp +  (1- v \overline{\xi'}) \overline{\xi'} } { (1-v \overline{\xi'})^2 - v^2 \overline{\epsilon'}_\perp \overline{\mu'}_\perp } , \label{eq:xiefflab}
\end{eqnarray}{}
which are Eqs. (24-28) in the main text.

\subsection{Plane wave dispersion relation}
With the above homogenised parameters, we have effective Maxwell's equations in the rest frame: 
\begin{eqnarray}
    \label{eq:Maxwellcomovhomog}
    \begin{bmatrix}
    0 & \nabla\times \\
    \nabla\times & 0
    \end{bmatrix} 
    \begin{bmatrix}
    \mathbf{E} \\
    \mathbf{H}
    \end{bmatrix} = \frac{\partial}{\partial t} 
    \begin{bmatrix}
    \hat{\boldsymbol{\epsilon}}^\text{eff} &  \hat{\boldsymbol{\xi}}^\text{eff} \\
    -\hat{\boldsymbol{\xi}}^\text{eff}& \hat{\boldsymbol{\mu}}^\text{eff}
    \end{bmatrix} 
    \begin{bmatrix}
    \mathbf{E} \\
    \mathbf{H}
    \end{bmatrix}
\end{eqnarray}
Assuming plane wave solutions, $\{\mathbf{E}(x,t),\mathbf{H}(x,t)\} = \{\mathbf{E}_0,\mathbf{H}_0\} e^{i(k_x+k_yy-\omega t)}$, we have
\begin{eqnarray}
    ik \begin{bmatrix} 
    E_z \\ H_y
    \end{bmatrix} &=& -i\omega \begin{bmatrix}
    \xi^\text{eff} & \mu^\text{eff}_\perp \\
    \epsilon^\text{eff}_\perp - \frac{k_y^2}{\omega^2 \mu^\text{eff}_{||}} & \xi^\text{eff}
    \end{bmatrix}\begin{bmatrix} 
    E_z \\ H_y
    \end{bmatrix}, \, \text{s-polarisation,} \\ 
    ik \begin{bmatrix} 
    E_y \\ H_z
    \end{bmatrix} &=& -i\omega \begin{bmatrix}
    \xi^\text{eff} & -\mu^\text{eff}_\perp + \frac{k_y^2}{\omega^2 \epsilon^\text{eff}_{||}} \\
    -\epsilon^\text{eff}_\perp  & \xi^\text{eff}
    \end{bmatrix}\begin{bmatrix} 
    E_y \\ H_z
    \end{bmatrix}, \, \text{p-polarisation.}
\end{eqnarray}
From which we can write dispersion relations as,
\begin{eqnarray}
    \mu^\text{eff}_\perp  \epsilon^\text{eff}_\perp \omega^2 &=& \frac{ \mu^\text{eff}_\perp}{  \mu^\text{eff}_{||}} k_y^2 + (k+ \xi^\text{eff}\omega)^2, \, \text{ s-polarisation,} \\
   \mu^\text{eff}_\perp  \epsilon^\text{eff}_\perp \omega^2 &=& \frac{  \epsilon^\text{eff}_\perp}{  \epsilon^\text{eff}_{||}} k_y^2 + (k+ \xi^\text{eff}\omega)^2, \, \text{ p-polarisation.}
\end{eqnarray}
At normal incidence we have, 
\begin{eqnarray}
    k = \omega\left(\pm \sqrt{\epsilon^\text{eff}_\perp\mu^\text{eff}_\perp } - \xi^\text{eff}\right),
\end{eqnarray}
for both polarisations.

\subsection{Modulation of only one parameter}
If only one parameter is modulated, while the effective material in the co-moving medium will in general still be bianisotropic ($\xi'\neq0$), in the lab frame the magnetoelectric coupling cancels out, and as discussed in the main text the low frequency response is reciprocal. It can be seen that $\xi^\text{eff}=0$ from Eq. \ref{eq:xiefflab}. Assuming only $\epsilon$ is modulated,
\begin{eqnarray}
    \xi^\text{eff} &\propto& \overline{\epsilon'}_{\perp}\overline{\mu'}_{\perp}v + (1-v\overline{\xi'})\overline{\xi'} \nonumber \\
    &=& \left< \frac{\epsilon(x')}{1-v^2/c(x')^2}  \right> \mu \left< \frac{1}{1-v^2/c(x')^2}  \right> v \nonumber \\ 
    &+& \left[1-v\left(-v\left< \frac{\epsilon(x')\mu}{1-v^2/c(x')^2} \right> \right)\right] \left(-v\left< \frac{\epsilon(x')\mu}{1-v^2/c(x')^2} \right> \right)  \nonumber \\
    &=& v\left< \frac{\epsilon(x')\mu}{1-v^2/c(x')^2}  \right> \left[\left< \frac{1}{1-v^2/c(x')^2}  \right> - 1 -v^2\left< \frac{\epsilon(x')\mu}{1-v^2/c(x')^2} \right> \right]  \nonumber \\
    &=&  v\left< \frac{1/c(x')^2}{1-v^2/c(x')^2}  \right> \left[ \left<  \frac{1 - 1 + v^2/c(x')^2 - v^2/c(x')^2}{1-v^2/c(x')^2}  \right> \right] = 0 
\end{eqnarray}{}
where $\left<\dots\right>$ stands for averaging in one unit cell in the co-moving frame.

\subsection{Maxwell's equations in the co-moving frame for each polarisation}
Here we explicitly write Maxwell's equations for each polarisation in the co-moving frame using a Galilean transformation, in an equivalent procedure to that detailed in Section \ref{sec:homogcomoving}. 

We start from Maxwell's equations in the lab frame, 
\begin{eqnarray}
    \nabla\times\mathbf{E} &=& -\frac{\partial \mathbf{B}}{\partial t}=-\frac{\partial }{\partial t} \left[ \mu(x,t) \mathbf{H}\right], \\
    \nabla\times\mathbf{H} &=& \frac{\partial \mathbf{D}}{\partial t}= \frac{\partial}{\partial t}  \left[ \epsilon(x,t)  \mathbf{E}\right].
\end{eqnarray}
We first look at $s$ polarization, $(E_z,H_x,H_y)$,
\begin{eqnarray}
    \frac{\partial E_z}{\partial y} &=& -\frac{\partial }{\partial t} \left[ \mu(x,t) H_x\right], \\
    \frac{\partial E_z}{\partial x}  &=& \frac{\partial }{\partial t} \left[ \mu(x,t) H_y\right],\\
    \frac{\partial H_y}{\partial x} - \frac{\partial H_x}{\partial y} &=& \frac{\partial}{\partial t}  \left[ \epsilon(x,t)  E_z\right],
\end{eqnarray}
Consider a moving coordinate frame of Galilean type,
\begin{eqnarray}
    x' &=& x-\Omega/g t = x-v t, \\
    t' &=& t.
\end{eqnarray}
where the parameters are only a function of the spatial coordinate, $ \epsilon(x')$ , $\mu(x')$.  We have,
\begin{eqnarray}
    \frac{\partial}{\partial x} &=& \frac{\partial}{\partial x'} \frac{\partial x'}{\partial x}  + \frac{\partial}{\partial t'} \frac{\partial t'}{\partial x}  = \frac{\partial}{\partial x'}, \\
    \frac{\partial}{\partial t} &=&  \frac{\partial}{\partial x'} \frac{\partial x'}{\partial t}  + \frac{\partial}{\partial t'} \frac{\partial t'}{\partial t} = \frac{\partial}{\partial t'} - v \frac{\partial}{\partial x'}, 
\end{eqnarray}
such that Maxwell's equations transform as,
\begin{eqnarray}
    \frac{\partial E_z}{\partial y'} &=& -\left( \frac{\partial}{\partial t'} - v \frac{\partial}{\partial x'} \right) \left[ \mu(x') H_x\right], \\
    \frac{\partial E_z}{\partial x'}  &=& \left( \frac{\partial}{\partial t'} - v \frac{\partial}{\partial x'} \right) \left[ \mu(x') H_y\right],\\
    \frac{\partial H_y}{\partial x'} - \frac{\partial H_x}{\partial y'} &=& \left( \frac{\partial}{\partial t'} - v \frac{\partial}{\partial x'} \right)  \left[ \epsilon(x')  E_z\right].
\end{eqnarray}
Rearranging, 
\begin{eqnarray} \label{eq:MaxwellMovingFrame1}
    && \frac{\partial E_z}{\partial y'} - v \frac{\partial}{\partial x'} \left[ \mu(x') H_x\right] = - \mu(x')  \frac{\partial H_x}{\partial t'}, \\ \label{eq:MaxwellMovingFrame2}
    && \frac{\partial}{\partial x'}\left[ E_z + v \mu(x') H_y\right] =  \mu(x') \frac{\partial H_y}{\partial t'},\\ \label{eq:MaxwellMovingFrame3}
   && \frac{\partial}{\partial x'} \left[H_y + v \epsilon(x')  E_z\right]  - \frac{\partial H_x}{\partial y'} =  \epsilon(x')  \frac{\partial E_z}{\partial t'} .
\end{eqnarray}
We use the change of variables, 
\begin{eqnarray}
    E_z' &=& E_z + v  \mu(x') H_y, \\
    H_y' &=& H_y + v \epsilon(x')  E_z,
\end{eqnarray}{}
and $ H'_x=H_x$. For the last two equations (\ref{eq:MaxwellMovingFrame2}, \ref{eq:MaxwellMovingFrame3}) we have, 
\begin{eqnarray} 
    && \frac{\partial E'_z}{\partial x'} =  \mu(x') \frac{\partial H_y}{\partial t'},\\    && \frac{\partial H'_y }{\partial x'}  - \frac{\partial H'_x}{\partial y'} =  \epsilon(x')  \frac{\partial E_z}{\partial t'} .
\end{eqnarray}
Using now the inverse change of variables, 
\begin{eqnarray}
    E_z &=&\frac{1}{1-v^2/c(x')^2} \left[ E'_z - v  \mu(x') H'_y \right], \label{eq:invvariab}\\
    H_y &=&\frac{1}{1-v^2/c(x')^2} \left[ H'_y - v  \epsilon(x')  E'_z \right],
\end{eqnarray}{}
where $c(x')=[\epsilon(x')\mu(x')]^{-1/2}$, we arrive at, 
\begin{eqnarray} 
     \frac{\partial E'_z}{\partial x'} &=& \frac{1}{1-v^2/c(x')^2} \left[  \mu(x')  \frac{ \partial H'_y}{\partial t'}      - v  \mu(x') \epsilon(x')  \frac{ \partial E'_z}{\partial t'}  \right],\\    
     \frac{\partial H'_y }{\partial x'}  - \frac{\partial H'_x}{\partial y'} &=& \frac{1}{1-v^2/c(x')^2} \left[ \epsilon(x')  \frac{\partial E'_z}{\partial t'}    - v  \epsilon(x')  \mu(x')  \frac{\partial H'_y}{\partial t'}  \right].
\end{eqnarray}
On the other hand, for the first equation in the set of Maxwell's equations in the co-moving frame (\ref{eq:MaxwellMovingFrame1}), using Eq. \ref{eq:invvariab}, 
\begin{eqnarray}
    \frac{1}{1-v^2/c(x')^2} \frac{\partial }{\partial y'}  \left[ E'_z - v  \mu(x') H'_y \right]- v \frac{\partial}{\partial x'} \left[ \mu(x') H_x'\right] = - \mu(x')  \frac{\partial H'_x}{\partial t'}
\end{eqnarray}{}
With a trivial operation, 
\begin{eqnarray}
    \frac{1}{1-v^2/c(x')^2} \frac{\partial }{\partial y'}  \left[ E'_z -\frac{v^2}{c(x')^2} E'_z  + \frac{v^2}{c(x')^2} E'_z  - v  \mu(x') H'_y \right]- v \frac{\partial}{\partial x'} \left[ \mu(x') H_x\right] = - \mu(x')  \frac{\partial H'_x}{\partial t'}
\end{eqnarray}{}
we can write, 
\begin{eqnarray}
    \frac{\partial E'_z }{\partial y'}  -  \frac{v }{1-v^2/c(x')^2} \frac{\partial }{\partial y'} \left[ \mu(x') H'_y - \frac{v}{c(x')^2} E'_z   \right]- v \frac{\partial}{\partial x'} \left[ \mu(x') H'_x\right] = - \mu(x')  \frac{\partial H_x}{\partial t'}
\end{eqnarray}{}
Taking, 
\begin{eqnarray}
    B'_x &=& \mu(x') H_x, \\
    B'_y &=&  \frac{1}{1-v^2/c(x')^2} \left[ \mu(x') H'_y - \frac{v}{c(x')^2} E'_z \right], 
\end{eqnarray}{}
we have, 
\begin{eqnarray}
    \frac{\partial E'_z}{\partial y'}   - v  \left[ \frac{\partial B'_y}{\partial y'} + \frac{\partial B'_x}{\partial x'}  \right]  = - \mu(x')  \frac{\partial H_x}{\partial t'}.
\end{eqnarray}{}
Finally, using $\nabla' \cdot \mathbf{B}' = 0$, we arrive at, 
\begin{eqnarray}
    \frac{\partial E'_z  }{\partial y'} = - \mu(x')  \frac{\partial H'_x}{\partial t'}.
\end{eqnarray}{}

The final set of equations reads as, 
\begin{eqnarray} 
    \frac{\partial E'_z}{\partial x'} &=& \frac{\mu(x')}{1-v^2 \epsilon(x') \mu(x')}  \frac{ \partial H'_y}{\partial t'}      - v  \frac{ \epsilon(x') \mu(x') }{1-v^2 \epsilon(x') \mu(x')}  \frac{ \partial E'_z}{\partial t'}  ,\\    
     \frac{\partial H'_y }{\partial x'}  - \frac{\partial H'_x}{\partial y'} &=& \frac{ \epsilon(x') }{1-v^2 \epsilon(x') \mu(x')}\frac{\partial E'_z}{\partial t'}    - v \frac{ \epsilon(x') \mu(x')}{1-v^2 \epsilon(x') \mu(x')}   \frac{\partial H'_y}{\partial t'} , \\
     \frac{\partial  E'_z}{\partial y'}  &=& - \mu(x')  \frac{\partial H'_x}{\partial t'},
\end{eqnarray}
which can be mapped to Maxwell's equations in a bianisotropic medium, 
\begin{eqnarray}
    \nabla'\times\mathbf{E}' &=& -\frac{\partial \mathbf{B'}}{\partial t'}=-\frac{\partial }{\partial t'} 
    \left[ \boldsymbol{\mu}' \mathbf{H'} + \boldsymbol{\zeta}' \mathbf{E'} \right], \\
    \nabla'\times\mathbf{H'} &=& \frac{\partial \mathbf{D'}}{\partial t'}= \frac{\partial}{\partial t'}  \left[ \boldsymbol{\epsilon}' \mathbf{E}' + \boldsymbol{\xi}'\mathbf{H'}\right],
\end{eqnarray}
\begin{eqnarray}
   \frac{\partial E'_z}{\partial x'} &=&  \mu_\perp' \frac{\partial H'_y }{\partial t'} - \zeta'_{yz} \frac{\partial E'_z }{\partial t'} ,\\    
    \frac{\partial H'_y }{\partial x'}  - \frac{\partial H'_x}{\partial y'} &=& \epsilon_\perp' \frac{\partial E'_z}{\partial t'} + \xi'_{zy} \frac{\partial H'_y}{\partial t'}  , \\
     \frac{\partial  E'_z}{\partial y'}  
    &=& - \mu_{||}'  \frac{\partial H'_x}{\partial t'},    
\end{eqnarray}
We see that Maxwell's equations are form-invariant under the chosen transformation, and that biansisotropic coupling emerges in the co-moving frame. 
From the above we can identify the relevant s-polarization effective medium parameters in the the co-moving frame, 
\begin{eqnarray}
     \overline{\mu'}_{||} &=& \left[ \frac{1}{d} \int_0^d\frac{1}{\mu'_{||}(x')}\,\text{d}x'\right]^{-1} ,\\
    \epsilon'_{\perp} &=& \frac{1}{d} \int_0^d  \frac{\epsilon(x')}{1-\epsilon(x')\mu(x') v^2} \text{d}x' , \\
    \mu'_{\perp} &=& \frac{1}{d} \int_0^d  \frac{\mu(x')}{1-\epsilon(x')\mu(x') v^2} \text{d}x' , \\ 
    \xi'_{zy}  &=& -\zeta'_{yz} = \xi' =   -v \frac{1}{d} \int_0^d  \frac{\epsilon(x')\mu(x')}{1-\epsilon(x')\mu(x') v^2} \text{d}x'
\end{eqnarray}

Similarly, the same derivation for $p$ polarisation, $(E_x,E_y,H_z)$, yields, 
\begin{eqnarray}
   \frac{\partial H'_z}{\partial x'} &=& - \frac{ \epsilon(x') }{1-v^2 \epsilon(x') \mu(x')} \frac{\partial E'_y }{\partial t'} - \frac{ \epsilon(x') \mu(x') }{1-v^2 \epsilon(x') \mu(x')} \frac{\partial H'_z }{\partial t'} ,\\    
    \frac{\partial E'_y }{\partial x'}  - \frac{\partial E'_x}{\partial y'} &=& -\frac{\mu(x') }{1-v^2 \epsilon(x') \mu(x')} \frac{\partial H'_z}{\partial t'} +  \frac{ \epsilon(x') \mu(x') }{1-v^2 \epsilon(x') \mu(x')} \frac{\partial E'_y}{\partial t'}  , \\
     \frac{\partial  H'_z}{\partial y'}  
    &=&  \frac{ \epsilon(x') }{1-v^2 \epsilon(x') \mu(x')}  \frac{\partial E'_x}{\partial t'},    
\end{eqnarray}
which map to
\begin{eqnarray}
   \frac{\partial H'_z}{\partial x'} &=& - \epsilon_\perp' \frac{\partial E'_y }{\partial t'} - \xi'_{yz} \frac{\partial H'_z }{\partial t'} ,\\    
    \frac{\partial E'_y }{\partial x'}  - \frac{\partial E'_x}{\partial y'} &=& -\mu_\perp' \frac{\partial H'_z}{\partial t'} - \zeta'_{zy} \frac{\partial E'_y}{\partial t'}  , \\
     \frac{\partial  H'_z}{\partial y'}  
    &=&  \epsilon_{||}'  \frac{\partial E'_x}{\partial t'},    
\end{eqnarray}
with parameters
\begin{eqnarray}
    \overline{\epsilon'}_{||} &=& \left[ \frac{1}{d} \int_0^d\frac{1}{\epsilon'_{||}(x')}\,\text{d}x'\right]^{-1}, \\ 
    \epsilon'_{\perp} &=& \frac{1}{d} \int_0^d  \frac{\epsilon(x')}{1-\epsilon(x')\mu(x') v^2} \text{d}x' , \\
    \mu'_{\perp} &=& \frac{1}{d} \int_0^d  \frac{\mu(x')}{1-\epsilon(x')\mu(x') v^2} \text{d}x' , \\ 
    \xi'_{yz}  &=& -\zeta'_{zy} = -\xi' =  v \frac{1}{d} \int_0^d  \frac{\epsilon(x')\mu(x')}{1-\epsilon(x')\mu(x') v^2} \text{d}x' 
\end{eqnarray}

In conclusion, combining the results for both polarisations and writing the effective material parameters in matrix form, we have in the co-moving frame, 
\begin{eqnarray}
    \hat{\boldsymbol{\epsilon}}' =\begin{bmatrix}
    \epsilon_{||}' & 0 & 0 \\
    0 & \epsilon_{\perp}' & 0 \\
    0 & 0 & \epsilon_{\perp}'
    \end{bmatrix}, \,
    \hat{\boldsymbol{\mu}}' =\begin{bmatrix}
    \mu_{||}' & 0 & 0 \\
    0 & \mu_{\perp}' & 0 \\
    0 & 0 & \mu_{\perp}'
    \end{bmatrix}, \,
    \hat{\boldsymbol{\xi}}' = -\boldsymbol{\zeta}' =\begin{bmatrix}
    0 & 0 & 0 \\
    0 & 0 & -\xi' \\
    0 & \xi' & 0
    \end{bmatrix},
\end{eqnarray}
with 
\begin{eqnarray}
    \xi'  &=& -v \frac{1}{d} \int_0^d  \frac{\epsilon(x')\mu(x')}{1-\epsilon(x')\mu(x') v^2} \text{d}x'.
\end{eqnarray}

\section{Exact theory in the absence of back-scattering}
\subsection{Exact theory based on an analytically-solvable first order PDE derived from Maxwell's equations}
Here we develop an exact theory valid in the absence of back-scattering. We start from Maxwell's equations in the lab frame,
\begin{eqnarray}
    \nabla\times\mathbf{E} &=& -\frac{\partial \mathbf{B}}{\partial t}, \\
    \nabla\times\mathbf{H} &=& \frac{\partial \mathbf{D}}{\partial t}.
\end{eqnarray}
Let us consider the components perpendicular to the space-time modulation, 
\begin{eqnarray}
    (\nabla\times\mathbf{E})_\perp &=& -\frac{\partial B_\perp}{\partial t}, \\
    (\nabla\times\mathbf{H})_\perp &=& \frac{\partial D_\perp}{\partial t}.
\end{eqnarray}
At normal incidence these reduce to,
\begin{eqnarray}
    \sigma_{s,p} \partial_x E_\perp = - \partial_t B_\perp, \\
    \sigma_{s,p} \partial_x H_\perp = - \partial_t D_\perp,
\end{eqnarray}
with and $\sigma_s=-1$ and $\sigma_p=+1$ for \textit{s-} and \textit{p-}polarisation, respectively. 

Let us now assume that the medium is impedance-matched at all positions and times, with equal permittivity and permeability modulations $\delta\epsilon(x,t)=\delta\mu(x,t)$, such that the medium has constant impedance,
\begin{eqnarray}
    Z=\sqrt{\frac{\mu_m\mu_0\,\delta\mu(x,t)}{\epsilon_m\epsilon_0\,\delta\epsilon(x,t)}} = \sqrt{\frac{\mu_m\mu_0}{\epsilon_m\epsilon_0}} = Z_m Z_0,
\end{eqnarray}
Thus, we can write, 
\begin{eqnarray}
    E_\perp &=&  \pm \sigma_{s,p} Z  H_\perp, \\
    D_\perp &=&  \pm  \sigma_{s,p} Z^{-1}  B_\perp,
\end{eqnarray}
with the top (bottom) sign corresponding to forward (backward) propagating waves. Hence, we can rewrite Maxwell's equations as, 
\begin{eqnarray}
    \sigma_{s,p} \partial_x \left[ \epsilon(x,t)^{-1} D_\perp \right] &=& - \partial_t (\pm\sigma_{s,p} Z  D_\perp), \\
    \sigma_{s,p}  \partial_x \left[\pm  \sigma_{s,p}  Z^{-1} \epsilon(x,t)^{-1} D_\perp \right] &=& - \partial_t D_\perp.
\end{eqnarray}
Since $Z$ is constant, we see that Maxwell's equations reduce to a single equation. We have, for both polarisations, 
\begin{eqnarray}
  \frac{\partial}{\partial x} \left[ \epsilon(x,t)^{-1} D_\perp \right] &=& \mp Z \frac{\partial D_\perp}{\partial t}  ,
\end{eqnarray}
with the $-$ and $+$ signs corresponding to forward and backward wave propagation, respectively. This is Eq. 35 in the main text. 

Next, we make a Galilean transformation to the co-moving frame, $x'=x-vt$, yielding, 
\begin{eqnarray}
  \mp \frac{\partial}{\partial x'} \left[ \epsilon(x')^{-1} D_\perp \right] &=&  Z \frac{\partial D_\perp}{\partial t'} - Z v \frac{\partial D_\perp}{\partial x'} ,
\end{eqnarray}
where we have made use of the fact that $\partial D_\perp'=\partial D_\perp$ for a Galilean transformation. Rearranging, 
\begin{eqnarray}
  \frac{\partial}{\partial x'} \left[ \left( \pm Z^{-1}\epsilon(x')^{-1} -  v \right) D_\perp \right] &=& -  \frac{\partial D_\perp}{\partial t'}.
\end{eqnarray}
In the frequency domain, 
\begin{eqnarray}
  \frac{\partial}{\partial x'} \left[ \left( \pm c(x') -  v \right) D_\perp \right] &=& i\omega  D_\perp.
\end{eqnarray}
where we have identified 
\begin{eqnarray}
    c(x') &=& \frac{1}{Z\epsilon(x')} = \frac{1}{\sqrt{\frac{\mu_m\mu_0}{\epsilon_m\epsilon_0}}\sqrt{[\epsilon_m\epsilon_0\delta\epsilon(x')]^2}} = \frac{c_mc_0}{\sqrt{\delta\epsilon(x')\delta\mu(x')}}  =\frac{1}{\sqrt{\epsilon(x')\mu(x')}}
\end{eqnarray}
as the local wave velocity, using that $\delta\epsilon(x')=\delta\mu(x') $. Operating, 
\begin{eqnarray}
  \frac{\partial \left[ \pm c(x') -  v  \right]}{\partial x'}  D_\perp + \left[  \pm c(x') -  v  \right] \frac{\partial D_\perp}{\partial x'}   &=& i\omega  D_\perp, \\
  \frac{1}{  \pm c(x') -  v  }\frac{\partial \left[ \pm c(x') -  v  \right]}{\partial x'} + \frac{1}{D_\perp}\frac{\partial D_\perp}{\partial x'}   &=& i\omega  \frac{1}{  \pm c(x') -  v  }, \\
  \frac{1}{  \pm c(x') -  v  }\frac{\partial \left[ \pm c(x') -  v  \right]}{\partial x'} + \frac{1}{D_\perp}\frac{\partial D_\perp}{\partial x'}   &=& i\omega  \frac{1}{  \pm c(x') -  v  }, \\
    \frac{\partial }{\partial x'} \left[ \ln{\left( \pm c(x') -  v  \right)} + \ln{ D_\perp} \right]   &=& i\omega  \frac{1}{  \pm c(x') -  v  }.
\end{eqnarray}
The above partial differential equation yields, 
\begin{eqnarray}
  \ln{\left[ D_\perp\left( \pm c(x') -  v  \right) \right]} = i\omega \int  \frac{1}{  \pm c(x') -  v  } \text{d}x'
\end{eqnarray}
This integral can be solved analytically for either travelling-wave stratified or sinusoidal space-time modulations, as we further detail in the next section. Let us here concentrate on the form of the solution, 
\begin{eqnarray}
  D_\perp = \frac{1}{  \pm c(x') -  v  } \exp{\left[i\omega  \int \frac{1}{  \pm c(x') -  v  } \text{d}x' \right]}.
\end{eqnarray}
The right hand side is purely imaginary in the cases considered in this work, $|c(x')|\leq v$, such that we can identify an effective wave-vector by considering that the phase change in the unit cell is $\Delta\phi = k_\text{eff}^{\pm} 2\pi/g$, with $g=2\pi/d$ the reciprocal lattice vector. Hence, 
\begin{eqnarray}
\label{eq:effkpm}
  k_\text{eff}^{\pm} = \omega\frac{g}{2\pi} \int_0^d \frac{1}{  \pm c(x') -  v  } \text{d}x'.
\end{eqnarray}

On the other hand, in the co-moving frame Maxwell's equations are of bianisotropic form,
\begin{eqnarray}
  \partial_x' \begin{bmatrix}
  E_\perp \\ H_\perp 
  \end{bmatrix} = \partial_t' 
  \begin{bmatrix}
    \xi' & -\sigma_{s,p}\mu'_\perp \\ -\sigma_{s,p}\epsilon_\perp' & \xi' 
  \end{bmatrix},
\end{eqnarray}
we assume a plane wave expansion, 
\begin{eqnarray}
  \pm i k'^{\pm} \begin{bmatrix}
  E_\perp \\ H_\perp 
  \end{bmatrix} =  i\omega' \begin{bmatrix}
    \xi' & -\sigma_{s,p}\mu'_\perp \\ -\sigma_{s,p}\epsilon'_\perp & \xi' 
  \end{bmatrix},
\end{eqnarray}
 and write the dispersion relation of forward and backward propagating waves, 
\begin{eqnarray}
  k'^{\pm} = \omega' \left(  \pm\sqrt{\epsilon'_\perp\mu'_\perp} - \xi' \right).
\end{eqnarray}
We have, 
\begin{eqnarray}
  k'^+  + k'^- &=& -2\omega' \xi', \\
  k'^+  - k'^- &=& 2\omega' \sqrt{\epsilon'_\perp\mu'_\perp}
\end{eqnarray}
From the above and using Eq. \ref{eq:effkpm} we have, 
\begin{eqnarray}
    \epsilon'_\perp &=& \mu'_\perp = \sqrt{\epsilon'_\perp\mu'_\perp}=  \frac{1}{2\omega'} \frac{\omega'}{d}\int_0^d \left[ \frac{1}{ c(x') - v} - \frac{1}{  -c(x') -  v  } \right] \text{d}x' \nonumber\\ && = \frac{1}{d} \int_0^d \frac{\epsilon(x')}{1-\epsilon(x')\mu(x')v^2} \text{d}x',\\
    \xi' &=& -\frac{1}{2\omega'} \frac{\omega'}{d}\int_0^d \left[ \frac{1}{ c(x') - v} + \frac{1}{  -c(x') -  v  } \right] \text{d}x' \nonumber\\ 
    && = - v \frac{1}{d} \int_0^d \frac{\epsilon(x')\mu(x')}{1-\epsilon(x')\mu(x')v^2} \text{d}x',
\end{eqnarray}
in agreement with our homogenisation formulae. This proves that the homogenisation theory is in fact exact when back-scattering is negligible, that is, if the medium is impedance-matched or, if it is not, whenever the frequency is low enough to be far away from  band gaps. 

\subsection{Exact transmission matrix theory for matched space-time media.}
Here we present an alternative proof based on transmission matrix theory that the homogenisation theory is exact at all frequencies for matched space-time media. 

We start from Maxwell's equations in the co-moving frame, which as derived above, can be written as
\begin{eqnarray}
    \begin{bmatrix}
    0 & \nabla'\times \\
    \nabla'\times & 0
    \end{bmatrix} 
    \begin{bmatrix}
    \mathbf{E'} \\
    \mathbf{H'}
    \end{bmatrix} = \frac{\partial}{\partial t'} 
    \begin{bmatrix}
    \hat{\boldsymbol{\epsilon}}'(x') &  \hat{\boldsymbol{\xi}}'(x') \\
    -\hat{\boldsymbol{\xi}}'(x')& \hat{\boldsymbol{\mu}}'(x')
    \end{bmatrix} 
    \begin{bmatrix}
    \mathbf{E'} \\
    \mathbf{H'}
    \end{bmatrix}
\end{eqnarray}
Assuming a plane wave expansion, $e^{ik'x'-i\omega't'}$, in a small enough region such that it is homogeneous, we can derive the normal incidence dispersion relation in the co-moving frame as, 
\begin{eqnarray}
    k^{\pm} &=& \omega'\left(\pm \sqrt{\epsilon'_\perp(x')\mu'_\perp(x') } - \xi'(x')\right) \nonumber \\
    &=& \omega'\left(\pm \frac{\sqrt{\epsilon(x')\mu(x') }}{1-\epsilon(x')\mu(x')v^2} + v\frac{\epsilon(x')\mu(x') }{1-\epsilon(x')\mu(x')v^2}\right) \nonumber  \\
    &=&  \omega'\frac{\pm\sqrt{\epsilon(x')\mu(x') }+v\epsilon(x')\mu(x')}{1-\epsilon(x')\mu(x')v^2}.
\end{eqnarray}
Operating, 
\begin{eqnarray}
    \label{eq:dispcomov}
    k^{\pm} &=& \pm \frac{\omega'}{c(x') \mp v}, 
\end{eqnarray}
with $c(x')=[\epsilon(x')\mu(x')]^{-1/2}$ the local wave velocity. We also have, 
\begin{eqnarray}
    \frac{E_\perp' }{H'_\perp} = \pm \sigma_{p,s} Z,
\end{eqnarray}
with constant impedance $Z=\sqrt{\mu(x')/\epsilon(x')}$. 

If the system is matched, waves propagate without reflections even through inhomogeneities, and the dispersion relation of Bloch modes can be found by enforcing that the phase accumulated in one spatial period is identical to the Bloch phase shift. This condition is, 
\begin{eqnarray}
    k' d = \int_0^d \text{d}x' k^{\pm} (x') +2\pi n, \,\, n=0,\pm1,\pm2,\cdots
\end{eqnarray}
where $d$ is the spatial period. Using Eq. \ref{eq:dispcomov}, and noting that under a Galilean transformation $k'=k$, we arrive at 
\begin{eqnarray} 
    k'^\pm = \pm\frac{1}{d}\int_0^d \text{d}x'  \frac{\omega'}{c(x') \mp v} +2\pi n, \,\, n=0,\pm1,\pm2,\cdots
\end{eqnarray}
This is in agreement with Eq. 38 in main text, which was derived by analytically solving Maxwell's equations and where we particularised to the fundamental Bloch band ($n=0$).

Alternatively, it is possible to write a transmission matrix between two points distanced by $\delta x' = d$, 
\begin{eqnarray}
    \mathbf{M} =  \mathbf{M}_0 e^{-i\Delta\phi},
\end{eqnarray}
where $\Delta\phi=(k^++k^-)d/2$ and $\mathbf{M}_0$ is the standard transmission matrix with wavenumber $\beta = (k^+ - k^-)/2$ and impedance $Z$, 
\begin{eqnarray}
    \mathbf{M}_0= \begin{bmatrix}
    \cos(\beta d) & -i Z \sin(\beta d) \\
    -i \sin(\beta d)/Z &  \cos(\beta d) 
    \end{bmatrix},
\end{eqnarray}
with $\Delta\phi=\omega'd\,\frac{v}{c^2-v^2}$ and $\beta = (k^+ - k^-)/2=\omega' \frac{c}{c^2-v^2}$. 

For two consecutive layers with matched impedance and infinitesimal thicknesses, 
\begin{eqnarray}
    \mathbf{M}_0 &=&  \begin{bmatrix}
    \cos(\beta_1 \delta_1) & -i Z \sin(\beta_1 \delta_1) \\
    -i \sin(\beta_1 \delta_1)/Z &  \cos(\beta_1 \delta_1) 
    \end{bmatrix} \cdot \begin{bmatrix}
    \cos(\beta_2 \delta_2) & -i Z \sin(\beta_2 \delta_2) \\
    -i \sin(\beta_2 \delta_2)/Z &  \cos(\beta_2 \delta_2) 
    \end{bmatrix}  \\
    &=& \begin{bmatrix}
    \cos(\beta_1 \delta_1 + \beta_2 \delta_2) & -i Z \sin(\beta_1 \delta_1 + \beta_2 \delta_2) \\
    -i \sin(\beta_1 \delta_1 + \beta_2 \delta_2)/Z &  \cos(\beta_1 \delta_1 + \beta_2 \delta_2) 
    \end{bmatrix},
\end{eqnarray}
and 
\begin{eqnarray}
    \Delta\phi = (k^+ + k^-) \frac{\delta_1}{2} + (k^+ + k^-) \frac{\delta_2}{2}
\end{eqnarray}

Generalising to an infinite number of layers, 
\begin{eqnarray}
    \mathbf{M}= \begin{bmatrix}
    \cos\left(\int \beta(x') \text{d}x'\right) & -i Z \sin\left(\int \beta(x') \text{d}x'\right) \\
    -i \sin\left(\int \beta(x') \text{d}x'\right)/Z &  \cos\left(\int \beta(x') \text{d}x'\right) 
    \end{bmatrix} \exp\left(-i \int \text{d}x' \frac{k^+ + k^-}{2}\right).
\end{eqnarray}
With the above transmission matrix the dispersion relation can be found from
\begin{eqnarray}
    \det\left(\mathbf{M} - \mathbbm{1} e^{-ik' d} \right)=0,
\end{eqnarray}
with $d$ the unit cell length. This leads to
\begin{eqnarray}
    \cos\left( k' d - \int \text{d}x' \frac{k^+ + k^-}{2}\right) = \cos\left( \int \frac{k^+ - k^-}{2} \text{d}x' \right), 
\end{eqnarray}
or, equivalently, 
\begin{eqnarray}
    \label{eq:dispersionrelatcomov}
    k'^\pm = \pm\frac{1}{d}\int_0^d \text{d}x'  \frac{\omega'}{c(x') \mp v} +2\pi n, \,\, n=0,\pm1,\pm2,\cdots
\end{eqnarray}
such that this is an alternative proof of Eq. (38) in the main text. 

The above expression constitutes the dispersion relation in the co-moving frame, 
\begin{eqnarray}
    k'^{\pm} = \frac{1}{v_\text{eff} } \omega' + \frac{2\pi}{d}n
\end{eqnarray}
which is then transformed to the lab frame through a Galilean transformation, $\omega' = \omega - v\,k$ and $k'=k$, 
\begin{eqnarray}
    \label{eq:dispersionrelatrest}
    \omega=  (v_\text{eff}+v) k'^{\pm} +  v_\text{eff} \frac{2\pi}{d}n.
\end{eqnarray}

\section{Stratified media.}

\subsection{Analytical effective parameters for bi-layer media}
Let us consider a travelling two-layer stratified medium with relative parameters $(\epsilon_1,\mu_1)$ and $(\epsilon_2,\mu_2)$, and thicknesses $d_1$ and $d_2$ (period $d=d_1+d_2$) moving at speed $v$. In this case, as discussed in the main text we have in the co-moving frame, 
\begin{eqnarray}
    \label{eq:epsefflab}
    \frac{\overline{\epsilon}'_{||}}{\epsilon_0} &=& d (\frac{d_1}{\epsilon_1}+\frac{d_2}{\epsilon_2})^{-1}, \\ 
     \frac{\overline{\mu}'_{||}}{\mu_0} &=& d (\frac{d_1}{\mu_1}+\frac{d_2}{\mu_2})^{-1}, \\ 
   \frac{\overline{ \epsilon}'_{\perp}}{\mu_0} &=& \frac{1}{d} \left(\frac{\epsilon_1}{1-\epsilon_1\mu_1v^2/c_0^2}d_1 +\frac{\epsilon_2}{1-\epsilon_2\mu_2v^2/c_0^2}d_2\right)  , \\ 
    \frac{\overline{\mu}'_{\perp}}{\mu_0} &=& \frac{1}{d} \left(\frac{\mu_1}{1-\epsilon_1\mu_1v^2/c_0^2}d_1 +\frac{\mu_2}{1-\epsilon_2\mu_2v^2/c_0^2}d_2\right) , \\ 
    \label{eq:xicomov}
    \overline{\xi'}c_0 &=& - \frac{v}{dc_0} \left(\frac{\epsilon_1\mu_1}{1-\epsilon_1\mu_1v^2/c_0^2}d_1 +\frac{\epsilon_2\mu_2}{1-\epsilon_2\mu_2v^2/c_0^2}d_2\right).
\end{eqnarray}{}
The above parameters have removable singularities at $v/c_0=1/\sqrt{\epsilon_1\mu_1}$ and $v/c_0=1/\sqrt{\epsilon_2\mu_2}$. Transforming to the rest frame, 
\begin{eqnarray}
    \frac{\epsilon^\text{eff}_{||} }{\epsilon_0}&=&  d (\frac{d_1}{\epsilon_1}+\frac{d_2}{\epsilon_2})^{-1}, 
    \\ 
     \frac{\mu^\text{eff}_{||} }{\mu_0}&=&  d (\frac{d_1}{\mu_1}+\frac{d_2}{\mu_2})^{-1}, 
    \\ 
 \frac{\epsilon^\text{eff}_{\perp} }{\epsilon_0}&=&    \frac{d \left(d_1 \epsilon_1 \left( 1- \epsilon_2 \mu_2  v^2 / c_0^2 \right)+d_2 \epsilon_2 \left( 1-\epsilon_1 \mu_1  v^2 / c_0^2\right)\right)}{d_1^2 \left( 1-\epsilon_2 \mu_2  v^2 / c_0^2\right)+d_1 d_2 \left( 2- (\epsilon_1 \mu_2  + \epsilon_2 \mu_1 )v^2/ c_0^2\right)+d_2^2 \left(1- \epsilon_1 \mu_1  v^2 / c_0^2\right)} 
 \\ 
 \frac{\mu^\text{eff}_{\perp} }{\mu_0}&=&    \frac{d \left(d_1 \mu_1 \left( 1- \epsilon_2 \mu_2  v^2 / c_0^2 \right)+d_2 \mu_2 \left( 1-\epsilon_1 \mu_1  v^2 / c_0^21\right)\right)}{d_1^2 \left( 1-\epsilon_2 \mu_2  v^2 / c_0^2\right)+d_1 d_2 \left( 2- (\epsilon_1 \mu_2  + \epsilon_2 \mu_1 )v^2/ c_0^2\right)+d_2^2 \left(1- \epsilon_1 \mu_1  v^2 / c_0^2\right)}
  \\ 
\xi^\text{eff}c_0 &=&    \frac{v d_1d_2(\epsilon_1-\epsilon_2)(\mu_1-\mu_2)/c_0}{d_1^2 \left( 1-\epsilon_2 \mu_2  v^2 / c_0^2\right)+d_1 d_2 \left( 2- (\epsilon_1 \mu_2  + \epsilon_2 \mu_1 )v^2/ c_0^2\right)+d_2^2 \left(1- \epsilon_1 \mu_1  v^2 / c_0^2\right)}.
\end{eqnarray}

In the absence of temporal modulation, $v=0$, the effective parameters reduce to, 
\begin{eqnarray}
    \frac{\epsilon^\text{eff}_{||} }{\epsilon_0}&=&  d (\frac{d_1}{\epsilon_1}+\frac{d_2}{\epsilon_2})^{-1}, 
    \\ 
     \frac{\mu^\text{eff}_{||} }{\mu_0}&=&  d (\frac{d_1}{\mu_1}+\frac{d_2}{\mu_2})^{-1}, 
    \\ 
 \frac{\epsilon^\text{eff}_{\perp} }{\epsilon_0}&=&    \frac{d \left(d_1 \epsilon_1 +d_2 \epsilon_2 \right)}{(d_1+d_2)^2 } =\frac{ d_1 \epsilon_1 +d_2 \epsilon_2 }{d }
 \\ 
 \frac{\mu^\text{eff}_{\perp} }{\mu_0}&=&    \frac{d \left(d_1 \mu_1 +d_2 \mu_2 \right)}{(d_1+d_2)^2 } =\frac{ d_1 \mu_1 +d_2 \mu_2 }{d }
  \\ 
\xi^\text{eff}&=&   0
\end{eqnarray}
consistent with the conventional homogenisation parameters of isotropic, non-magneto-electric, bi-layer crystals. 
For temporal-only modulations, $v\rightarrow\infty$,
\begin{eqnarray}
    \frac{\epsilon^\text{eff}_{||} }{\epsilon_0}&=&  d (\frac{d_1}{\epsilon_1}+\frac{d_2}{\epsilon_2})^{-1}, 
    \\ 
     \frac{\mu^\text{eff}_{||} }{\mu_0}&=&  d (\frac{d_1}{\mu_1}+\frac{d_2}{\mu_2})^{-1}, 
    \\ 
 \frac{\epsilon^\text{eff}_{\perp} }{\epsilon_0}&=&    d (\frac{d_1}{\epsilon_1}+\frac{d_2}{\epsilon_2})^{-1}, 
 \\ 
 \frac{\mu^\text{eff}_{\perp} }{\mu_0}&=&    d (\frac{d_1}{\mu_1}+\frac{d_2}{\mu_2})^{-1}, 
  \\ 
\xi^\text{eff} &=&   0,
\end{eqnarray}
as expected for time-like bilayer crystals. 

\subsection{Effective group velocity}
From Eq. \ref{eq:dispersionrelatcomov}, it is straightforward to arrive to,
\begin{eqnarray}
    k'^{\pm} = \omega' \left( \frac{d_1/d}{c_1c_0 - v} + \frac{d_2/d}{c_2c_0 - v} \right) + \frac{2\pi}{d}n
\end{eqnarray}
with $c_i=1/\sqrt{\epsilon_i\mu_i}$. Transforming to the rest frame through Eq. \ref{eq:dispersionrelatrest}, 
\begin{eqnarray}
    \omega = \pm\left[\left( \frac{d_1/d}{c_1c_0 - v} + \frac{d_2/d}{c_2c_0 - v} \right)^{-1} + v \right] k'^{\pm} + \left( \frac{d_1/d}{c_1c_0 - v} + \frac{d_2/d}{c_2c_0 - v} \right)^{-1} \frac{2\pi}{d}n
\end{eqnarray}
we identify the effective group velocity as, 
\begin{eqnarray} \label{eq:effvelstratgeneral}
    v_\text{eff}^{\pm} = \left( \frac{d_1/d}{\pm c_1c_0 - v} + \frac{d_2/d}{\pm c_2c_0 - v} \right)^{-1} + v.
\end{eqnarray}

\subsection{Symmetric modulations }
Considering symmetric modulations above and below a background medium, $\epsilon_{1,2}=\epsilon_m(1\pm\alpha_e)$, $\mu_{1,2}=\mu_m(1\pm\alpha_m)$, with $d_1=d_2$, the above expressions for effective parameters in the rest frame yield, 
\begin{eqnarray}
 \frac{\epsilon^\text{eff}_{\perp} }{\epsilon_0}&=&    \epsilon_m \frac{1-\left(1- \alpha_e ^2\right) v^2c_m^{-2}c_0^{-2}}{1-v^2c_m^{-2}c_0^{-2}}
 \\ 
 \frac{\mu^\text{eff}_{\perp} }{\epsilon_0}&=&    \mu_m \frac{1-\left(1- \alpha_m ^2\right) v^2c_m^{-2}c_0^{-2}}{1-v^2c_m^{-2}c_0^{-2}}
  \\ 
\xi^\text{eff}c_0 &=&  \alpha_e\alpha_m   \frac{v c_0^{-1}c_m^{-2}}{ 1-v^2c_m^{-2}c_0^{-2}},
\end{eqnarray}
from where we can obtain the effective wave velocities as, 
\begin{eqnarray}
    v_\text{eff}^+ &=& +c_0 c_m\frac{(v+c_mc_0)(-v+c_mc_0)}{v c_mc_0\alpha_e\alpha_m+\sqrt{\left[ c_m^2c_0^2 - v^2(1-\alpha_e^2)\right]\left[ c_m^2c_0^2 - v^2(1-\alpha_m^2)\right]}}, \\
    v_\text{eff}^- &=& -c_0 c_m\frac{(v+c_mc_0)(-v+c_mc_0)}{v c_mc_0\alpha_e\alpha_m-\sqrt{\left[ c_m^2c_0^2 - v^2(1-\alpha_e^2)\right]\left[ c_m^2c_0^2 - v^2(1-\alpha_m^2)\right]}}.
\end{eqnarray}
It is straightforward to see that the same result is obtained by calculating the dispersion in the co-moving frame and transforming to the rest frame, Eq. \ref{eq:effvelstratgeneral}. The above equation for the forward wave group velocity shows a pole for modulation speeds,
\begin{eqnarray}
    v_c = \frac{c_m c_0}{\sqrt{1-\alpha_e^2-\alpha_m^2+\alpha_e^2\alpha_m^2}}.
\end{eqnarray}
For equal-amplitude electric and magnetic modulations, $(\alpha_e,\alpha_m) = \sqrt{2}\alpha(\cos\phi,\sin\phi)$, 
\begin{eqnarray}
    v_c = \frac{c_m c_0}{\sqrt{(1-2\alpha^2\cos^2\phi) (1-2\alpha^2\sin^2\phi)}}.
\end{eqnarray}
From the above equation, we have that the critical modulation for only electric (or magnetic) modulations, $\phi=0,\, \phi$ ($\phi=\pi/2,\, 3\pi/2$), the pole is at,
\begin{eqnarray}
    v_c = \frac{c_m c_0}{\sqrt{1-2\alpha^2}}
\end{eqnarray}
while $|\alpha_e|=\pm|\alpha_m|$ ($\phi=\pm\pi/4,\, \pm3\pi/4 $), it is at,
\begin{eqnarray}
    v_c = \frac{c_m c_0}{1-\alpha^2}.
\end{eqnarray}
Hence to second order in $\alpha$ the pole does not depend on $\phi$.

On the other hand, for the matched bi-layer medium considered in the main text, with $\alpha_e=\alpha_m=\alpha$, the above expressions for the effective parameters reduce to,
\begin{eqnarray}
 \frac{\epsilon^\text{eff}_{\perp} }{\epsilon_0}&=&    \epsilon_m \frac{1-\left(1- \alpha ^2\right) v^2c_m^{-2}c_0^{-2}}{1-v^2c_m^{-2}c_0^{-2}}
 \\ 
 \frac{\mu^\text{eff}_{\perp} }{\epsilon_0}&=&    \mu_m \frac{1-\left(1- \alpha ^2\right) v^2c_m^{-2}c_0^{-2}}{1-v^2c_m^{-2}c_0^{-2}}
  \\ 
\xi^\text{eff}c_0 &=&  \alpha ^2   \frac{v c_0^{-1}c_m^{-2}}{ 1-v^2c_m^{-2}c_0^{-2}},
\end{eqnarray}
and effective wave velocities are, 
\begin{eqnarray}
    v_\text{eff}^+ &=& +c_0 c_m\frac{1-vc_m^{-1}c_0^{-1}}{1-vc_m^{-1}c_0^{-1}(1-\alpha^2)}, \\
    v_\text{eff}^- &=& -c_0 c_m\frac{1+vc_m^{-1}c_0^{-1}}{1+vc_m^{-1}c_0^{-1}(1-\alpha^2)},
\end{eqnarray}
which are Eq. (49) of the main text.

\section{Sinusoidal travelling-wave modulations}

\subsection{Analytical expressions for effective parameters}
In this case we need to evaluate the homogenisation integrals for parameters in the co-moving frame,
\begin{eqnarray}
    \epsilon(x') &=& \epsilon_m\epsilon_0[1 + 2\alpha_e\cos(gx')], \\
    \mu(x')      &=& \mu_m\mu_0[1 + 2\alpha_m\cos(gx-\Omega t)]. 
\end{eqnarray}
We have for the parallel components of permittivity and permeability, 
\begin{eqnarray}
    \overline{\epsilon}'_{||} &=& \left[ \frac{1}{d}         \int_0^d \frac{1}{\epsilon(x')}\,\text{d}x'\right]^{-1} = \epsilon_m\epsilon_0 \sqrt{(1-2\alpha_e)(1+2\alpha_e)}, 
    \\ \overline{\mu}'_{||} &=& \left[ \frac{1}{d}         \int_0^d \frac{1}{\mu(x')}\,\text{d}x'\right]^{-1} = \mu_m\mu_0 \sqrt{(1-2\alpha_m)(1+2\alpha_m)},
\end{eqnarray} 
For the perpendicular components and the magnetoelectric coupling,
\begin{eqnarray}
    \overline{\epsilon}'_{\perp} &=& \frac{1}{d} \int_0^d      \frac{\epsilon(x')}{1-\epsilon(x')\mu(x') v^2} \text{d}x' \nonumber
    \\ &=&  \frac{\epsilon_m\epsilon_0}{d} \int_0^d  \frac{1 + 2\alpha_e\cos(gx')}{1-[1 + 2\alpha_e\cos(gx')][1 + 2\alpha_m\cos(gx')] v^2c_m^{-2}c_0^{-2}} \text{d}x', 
    \\
    \overline{\mu}'_{\perp} &=& \frac{1}{d} \int_0^d      \frac{\mu(x')}{1-\epsilon(x')\mu(x') v^2} \text{d}x'  \nonumber
    \\ &=&  \frac{\mu_m\mu_0}{d} \int_0^d  \frac{1 + 2\alpha_m\cos(gx')}{1-[1 + 2\alpha_e\cos(gx')][1 + 2\alpha_m\cos(gx')] v^2c_m^{-2}c_0^{-2}} \text{d}x', 
    \\ 
    \overline{\xi}' &=&-  \frac{ v }{d} \int_0^d  \frac{\epsilon(x')\mu(x')}{1-\epsilon(x')\mu(x') v^2} \text{d}x' \nonumber \\ &=& - \frac{ v c_m^{-2}c_0^{-2}}{d} \int_0^d  \frac{[1 + 2\alpha_e\cos(gx')][1 + 2\alpha_m\cos(gx')]}{1-[1 + 2\alpha_e\cos(gx')][1 + 2\alpha_m\cos(gx')] v^2c_m^{-2}c_0^{-2}} \text{d}x' .
\end{eqnarray}
The above integrals can be solved analytically in the complex plane. 

For the permittivity, 
\begin{eqnarray}
    \overline{\epsilon}'_{\perp} &=&  \frac{\epsilon_m\epsilon_0}{2\pi} \int_0^{2\pi}  \frac{1 + 2\alpha_e\cos(gx')}{1-[1 + 2\alpha_e\cos(gx')][1 + 2\alpha_m\cos(gx')] v^2c_m^{-2}c_0^{-2}} \text{d}(gx') \nonumber \\ 
    &=& 2\frac{\epsilon_m\epsilon_0}{2\pi} \int_{-1}^{+1}  \frac{1 + 2\alpha_ey}{1-[1 + 2\alpha_ey][1 + 2\alpha_my] v^2c_m^{-2}c_0^{-2}} \frac{\text{d}y}{\sqrt{1-y^2}}
\end{eqnarray}
which has poles at 
\begin{equation}
    y_{\pm} = \frac{-(\alpha_e+\alpha_m)\pm\sqrt{(\alpha_e+\alpha_m)^2-4\alpha_e\alpha_m(1-v^{-2}c_m^2c_0^2)}}{4\alpha_e\alpha_m},
\end{equation}
with residues 
\begin{eqnarray}
    R_+^{\epsilon} &=& -\frac{(1+2\alpha_ey_+)v^{-2}c_m^2c_0^2}{4\alpha_e\alpha_my_++\alpha_e+\alpha_m}\frac{\text{sgn}(v-c_mc_0)}{\sqrt{y_+^2-1}}, \\
    R_-^{\epsilon} &=& -\frac{(1+2\alpha_ey_-)v^{-2}c_m^2c_0^2}{4\alpha_e\alpha_my_-+\alpha_e+\alpha_m}\frac{1}{\sqrt{y_-^2-1}},
\end{eqnarray}
and we have, 
\begin{eqnarray}
    \overline{\epsilon}'_{\perp} &=&  \frac{\epsilon_m\epsilon_0}{2} \left[R_+^{\epsilon} + R_-^{\epsilon} \right] .
\end{eqnarray}

Similarly, for the permeability, 
\begin{eqnarray}
    \overline{\mu}'_{\perp} &=&  \frac{\mu_m\mu_0}{2} \left[R_+^{\mu} + R_-^{\mu} \right], 
\end{eqnarray}
with, 
\begin{eqnarray}
    R_+^{\mu} &=& -\frac{(1+2\alpha_m y_+)v^{-2}c_m^2c_0^2}{4\alpha_e\alpha_my_++\alpha_e+\alpha_m}\frac{\text{sgn}(v-c_mc_0)}{\sqrt{y_+^2-1}}, \\
    R_-^{\mu} &=& -\frac{(1+2\alpha_m y_-)v^{-2}c_m^2c_0^2}{4\alpha_e\alpha_my_-+\alpha_e+\alpha_m}\frac{1}{\sqrt{y_-^2-1}}.
\end{eqnarray}
For the magneto-electric coupling,
\begin{eqnarray}
    \overline{\xi}'_{\perp} &=&  \frac{1}{2} \left[R_+^{\xi} + R_-^{\xi} \right], 
\end{eqnarray}
with, 
\begin{eqnarray}
    R_+^{\xi} &=& \frac{1}{v}\left[ 1+ \frac{v^{-2}c_m^2c_0^2}{4\alpha_e\alpha_my_++\alpha_e+\alpha_m}\frac{\text{sgn}(v-c_mc_0)}{\sqrt{y_+^2-1}}\right], \\
    R_-^{\xi} &=& \frac{1}{v}\left[ 1+ \frac{v^{-2}c_m^2c_0^2}{4\alpha_e\alpha_my_-+\alpha_e+\alpha_m}\frac{1}{\sqrt{y_-^2-1}}\right].
\end{eqnarray}
Finally, these homogenised parameters have to be transformed to the rest frame with Eqs. (24-28) in the main text (section IA of this S.M.).  

\subsection{Results for non-matched systems}

\begin{figure}[b]
    \centering
    \includegraphics[width=0.25\columnwidth]{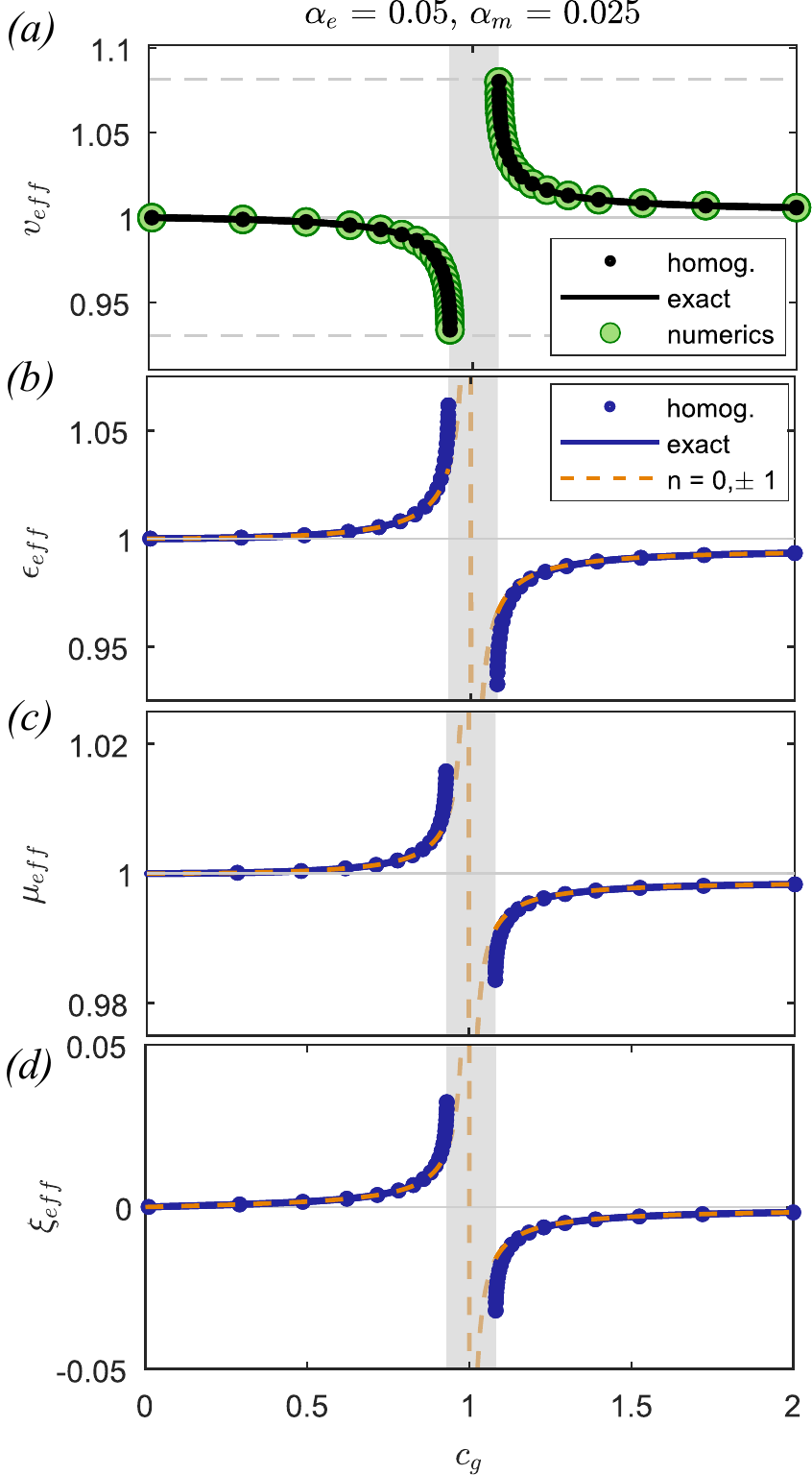}
    \caption{Same as Fig. 2 of the main text but when  space-time modulations of $\epsilon$ and $\mu$ are not matched, with $\alpha_e=0.05$, $\alpha_m=0.025$, and $\epsilon_m=\mu_m=1$. }
    \label{figSM}
\end{figure}

Figure \ref{figSM} is the same as Fig. 3 of the main text but for a non-matched system.

{}

\subsection{Fresnel drag and equivalent moving medium}

We start from the effective bianisotropic medium in the rest frame, 
\begin{eqnarray}
    \boldsymbol{\epsilon} = \left[ 
    \begin{array}{ccc} 
        \epsilon_{||} & 0 & 0 \\
        0 & \epsilon & 0 \\
        0 & 0 & \epsilon 
    \end{array} \right] ; \;
    \boldsymbol{\mu} = \left[ 
    \begin{array}{ccc} 
        \mu_{||} & 0 & 0 \\
        0 & \mu & 0 \\
        0 & 0 & \mu 
    \end{array} \right] ; \;
    \boldsymbol{\xi} = \boldsymbol{\zeta}^T =\left[ 
    \begin{array}{ccc} 
        0 & 0 & 0 \\
        0 & 0 & +\xi \\
        0 & -\xi & 0 
    \end{array} \right],  
\end{eqnarray}
where we have dropped the label $\text{eff}$ for conciseness, as well as $\perp$ in the perpendicular tensor components. 

The bianisotropic medium can be mapped to a medium moving at speed $v_D$ with (non-bianisotropic) parameters, 
\begin{eqnarray}
    \boldsymbol{\epsilon}_\text{eq} = \left[ 
    \begin{array}{ccc} 
        \epsilon_{\text{eq},||} & 0 & 0 \\
        0 & \epsilon_\text{eq} & 0 \\
        0 & 0 & \epsilon_\text{eq}
    \end{array} \right] ; \;
    \boldsymbol{\mu}_\text{eq} = \left[ 
    \begin{array}{ccc} 
        \mu_{\text{eq},||} & 0 & 0 \\
        0 & \mu_\text{eq} & 0 \\
        0 & 0 & \mu_\text{eq}
    \end{array} \right]. 
\end{eqnarray} 
From Ref. \cite{Kong}, in the $EH$ representation the constitutive parameters change as, 
\begin{align} 
    \label{eq:movingparameters}
    \epsilon &= \epsilon_\text{eq}\frac{1-v_D^2/c_0^2}{1-n_\text{eq}^2v_D^2/c_0^2} =  \epsilon_\text{eq}\frac{1-v_D^2/c_0^2}{1-\epsilon_\text{eq}\mu_\text{eq}v_D^2}, \, & \epsilon_{||}  =  \epsilon_{\text{eq},||},  \\
     \mu &= \mu_\text{eq}\frac{1-v_D^2/c_0^2}{1-n_\text{eq}^2v_D^2/c_0^2} =  \mu_\text{eq}\frac{1-v_D^2/c_0^2}{1-\epsilon_\text{eq}\mu_\text{eq}v_D^2},  \, &\mu_{||} = \mu_{\text{eq},||}, \\
    \xi &= -\frac{v_D}{c_0^2}\frac{n_\text{eq}^2-1}{1-n_\text{eq}^2v_D^2/c_0^2} = -v\frac{\epsilon_\text{eq}\mu_\text{eq}-1/c_0^2}{1-\epsilon_\text{eq}\mu_\text{eq}v_D^2}, & 
\end{align}
Noticing $\epsilon/\mu=\epsilon_\text{eq}/\mu_\text{eq}$, we have, 
\begin{eqnarray} 
    \epsilon &=& \epsilon_\text{eq}\frac{1-v_D^2/c_0^2}{1-\epsilon_\text{eq}^2\mu v_D^2/\epsilon} \\
    \xi &=& v_D\frac{\epsilon_\text{eq}^2\mu/\epsilon -1/c_0^2}{1-\epsilon_\text{eq}^2\mu v_D^2/\epsilon},
\end{eqnarray}
    Solving,
\begin{align}
    \epsilon_\text{eq} &=  \frac{1+c_0^2 (\epsilon\mu + \xi ^2) +\sqrt{\left(1 - c_0^2 \left(\epsilon\mu-\xi ^2\right) \right)^2-4 c_0^2 \mu  \epsilon }}{2 c_0^2\mu }, \label{eq:epseq} \\
   \mu_\text{eq} &= \frac{1+c_0^2 (\epsilon\mu - \xi ^2) +\sqrt{\left(1- c_0^2 \left(\epsilon\mu-\xi ^2\right) \right)^2-4 c_0^2 \mu  \epsilon }}{2 c^2\epsilon } , \label{eq:mueq} \\
    v_D &= \frac{-1+c_0^2 ( \epsilon\mu-\xi ^2) -\sqrt{\left(1+ c_0^2 \left(\epsilon\mu-\xi ^2\right) \right)^2-4 c_0^2 \mu  \epsilon }}{2 \xi }. \label{eq:veq}
\end{align}{}
Together with the exact effective medium parameters, we can use the above expressions to derive the exact parameters of the equivalent moving medium. For the matched case, using Eqs. (53-57) from the main text, we have, 
\begin{align}
    \frac{\epsilon_\text{eq}}{ \epsilon_0 \epsilon_m} &= \frac{\mu_\text{eq}}{\mu_0\mu_m} = \frac{c_0 c_m}{v (\Gamma_-\mp\Gamma_+)}  \\
     &
     \left[ \mp1 \pm \Gamma_+ +\Gamma_- +  \Gamma_+ \Gamma_- \left(-1 + \frac{v^2}{c_0^2}\right)  + \sqrt{\left(\Gamma_-^2  \left(1-\frac{v^2}{c_0^2}\right)  \pm 2 \Gamma_- -1\right) \left(\Gamma_+^2  \left(1-\frac{v^2}{c_0^2}\right) + 2 \Gamma_+ -1\right)}\right] , \nonumber \\
     v_D &= \frac{c_0^2}{v (\Gamma_- (2 \Gamma_+-1)\mp\Gamma_+)} \\
     & \left[\mp1 \pm \Gamma_+ +\Gamma_- -  \Gamma_+ \Gamma_-  \left(1 + \frac{v^2}{c_0^2}\right)+\sqrt{\left(\Gamma_-^2  \left(1-\frac{v^2}{c_0^2}\right)  \pm 2 \Gamma_- -1\right) \left(\Gamma_+^2  \left(1-\frac{v^2}{c_0^2}\right) + 2 \Gamma_+ -1\right)} \right], \nonumber
\end{align} 
where the top (bottom) sign corresponds to sub(super)-luminal modulations and with 
\begin{equation}
    \Gamma_{\pm} = \frac{c_mc_0}{\sqrt{(1-4\alpha^2) (v\pm v_c^+)(v\pm v_c^-)}}.
\end{equation}

On the other hand, we can use the perturbation theory parameters obtained from Floquet-Bloch theory assuming three modes only \cite{huidobro2019}, 
\begin{eqnarray}
    \epsilon_{||} &=&  \epsilon_m \label{eq:effparam1}\\
    \mu_{||} &=& \mu_m \\
    \epsilon &=& \epsilon_m \left( 1 + \alpha_e^2\frac{2\Omega^2}{c_m^2g^2-\Omega^2}\right) \\
    \mu &=&  \mu_m\left( 1 + \alpha_m^2\frac{2\Omega^2}{c_m^2g^2-\Omega^2} \right) \\
    \xi &=& \alpha_e\alpha_m\frac{2g\Omega}{c_m^2g^2-\Omega^2},  \label{eq:effparam2}
\end{eqnarray}
Assuming $\epsilon=\mu$ and keeping terms $\mathcal{O}(\alpha^2)$, we can write, 
\begin{eqnarray}
    \epsilon_\text{eq} &\approx& \epsilon_m \left( 1 + \alpha^2\frac{2\Omega^2}{c_m^2g^2-\Omega^2}\right) \label{eq:epsilonmovnew}\\
    v_D &\approx& - \alpha^2 \frac{1}{c_0^{-2}-c_m^{-2}} \left( \frac{2g\Omega}{c_m^2g^2-\Omega^2}\right)  \label{eq:velmovnew}
\end{eqnarray}{}

\bibliography{supp}